\documentclass[12pt]{article}
\usepackage[margin=0.8in]{geometry}

\usepackage{graphicx}  
\usepackage{dcolumn}   
\usepackage{bm}        
\usepackage{amssymb}   
\usepackage{amsmath}   
\usepackage{amsthm}    
\usepackage{xcolor}    
\usepackage{colortbl} 
\usepackage{tikz}      
\usepackage{ifthen}    
\usepackage{multirow}  
\usepackage{tabu}      
\usepackage{soul}      
\usepackage{subfigure}
\usepackage{xspace}
\usepackage{adjustbox}
\usepackage{etoolbox}
\usepackage{xfp}
\usepackage{authblk}
\usepackage[unicode=true,
  linktocpage,
  linkbordercolor={0.5 0.5 1},
  citebordercolor={0.5 1 0.5},
  linkcolor=blue]{hyperref}
\usepackage[numbers]{natbib}
\newcommand{\highTc}{high-$T_{\rm c}$\xspace}
\newcommand{\Tc}{$T_{\rm c}$\xspace}
\newcommand{\NdLSCO}{La$_{\rm 1.6-x}$Nd$_{\rm 0.4}$Sr$_x$CuO$_4$\xspace}

\newcommand{\ps}{\ensuremath{p^{\star}}\xspace}
\newcommand{\Ts}{$T^{\rm \star}$\xspace}
\newcommand{\degr}[1]{\ensuremath{#1^{\circ}}\xspace}
\newcommand{\Tlinear}{$\mathit{T}$-linear\xspace}

\definecolor{editblue}{rgb}{.4,.4,1}

\newcounter{para}

\begin{document}

\title{Linear-in temperature resistivity from an isotropic Planckian scattering rate}

\author[1,2,3]{Ga\"el Grissonnanche}
\author[2]{Yawen Fang}
\author[1,4]{Ana\"elle Legros}
\author[1]{Simon Verret}
\author[1]{Francis Lalibert\'e}
\author[1]{Cl\'ement Collignon}
\author[5]{Jianshi Zhou}
\author[6]{David Graf}
\author[7]{Paul A. Goddard}
\author[1,8]{Louis Taillefer \thanks{louis.taillefer@usherbrooke.ca}}
\author[2,8]{B.~J.~Ramshaw \thanks{bradramshaw@cornell.edu}}

\affil[1]{D\'epartement de physique, Institut quantique, and RQMP, Universit\'e de Sherbrooke, Sherbrooke, Qu\'ebec, Canada}
\affil[2]{Laboratory of Atomic and Solid State Physics, Cornell University, Ithaca, NY, USA}
\affil[3]{Kavli Institute at Cornell for Nanoscale Science, Ithaca, NY, USA}
\affil[4]{SPEC, CEA, CNRS-UMR 3680, Universit\'e Paris-Saclay, Gif-sur-Yvette, France }
\affil[5]{Materials Science and Engineering Program, Department of Mechanical Engineering, University of Texas at Austin, Austin, TX, USA}
\affil[6]{National High Magnetic Field Laboratory, FL, USA}
\affil[7]{Department of Physics, University of Warwick, Coventry, UK}
\affil[8]{Canadian Institute for Advanced Research, Toronto, Ontario, Canada}

\newpage
\date{}
\maketitle

\section*{}

\textbf{
A variety of ``strange metals'' exhibit resistivity that decreases linearly with temperature as $T\rightarrow 0$ \cite{martin1990,lohneysen1994,doiron2009correlation}, in contrast with conventional metals where resistivity decreases as $T^2$. This \Tlinear resistivity has been attributed to charge carriers scattering at a rate given by $\bm{\hbar/\tau=\alpha k_{\rm B} T}$, where $\bm{\alpha}$ is a constant of order unity. This simple relationship between the scattering rate and temperature is observed across a wide variety of materials, suggesting a fundamental upper limit on scattering---the ``Planckian limit'' \cite{bruin2013similarity,legros2019universal}---but little is known about the underlying origins of this limit. Here we report a measurement of the angle-dependent magnetoresistance (ADMR) of \NdLSCO---a hole-doped cuprate that displays \Tlinear resistivity down to the lowest measured temperatures \cite{daou2009linear}. The ADMR unveils a well-defined Fermi surface that agrees quantitatively with angle-resolved photoemission spectroscopy (ARPES) measurements \cite{matt2015electron} and reveals a $\bm{T}$-linear scattering rate that saturates the Planckian limit, namely $\bm{\alpha = 1.2 \pm 0.4}$. Remarkably, we find that this Planckian scattering rate is isotropic, i.e. it is independent of direction, in contrast with expectations from ``hot-spot'' models \cite{hlubina1995resistivity,stojkovic1997theory}. Our findings suggest that \Tlinear resistivity in strange metals emerges from a momentum-independent inelastic scattering rate that reaches the Planckian limit.
}

\section*{Introduction}

Immediately following the discovery of \highTc superconductivity in the cuprates, it was noted that the normal-state resistivity is linear over a broad temperature range \cite{gurvitch1987resistivity}. \Tlinear resistivity extending to low temperatures indicates a strongly correlated metallic state, and it was recognized early on that understanding \Tlinear resistivity may be the key to unraveling the mystery of \highTc superconductivity itself \cite{varma1989phenomenology}. Since then, \Tlinear resistivity has become a widespread phenomenon in strongly correlated metals, occurring in systems as diverse as organic and iron-based superconductors \cite{doiron2009correlation} and magic angle twisted bilayer graphene \cite{cao_matbg2020}. The fact that \Tlinear resistivity is often found in proximity to unconventional superconductivity is highly suggestive of a common underlying origin, but \Tlinear resistivity at low temperatures lies outside the standard Fermi-liquid description of metals and thus remains a central unsolved problem in quantum materials research.

The difficulty in developing a controlled, microscopic theory of \Tlinear resistivity has led to the creation of new theoretical approaches that draw on techniques developed for the study of quantum gravity, including holography and the Sachdev-Ye-Kitaev model \cite{parcollet1999non,davison2014holographic,hartnoll2015theory,patel2019theory,cha2020linear}. While these theories are not  microscopically motivated, they explicitly account for strong quasiparticle interactions in a controlled way and suggest that $T$-linear resistivity might emerge as a universal principle---independent of microscopic details. The transport scattering rate $1/\tau$ in these models obeys the so-called ``Planckian limit'':
\begin{equation}
\frac{\hbar}{\tau} = \alpha k_B T,
\label{eq:planck}
\end{equation}
where $k_B$ and $\hbar$ are the Boltzmann and Planck constants, respectively, and $\alpha$ is a constant of order unity. Simple estimates of $\alpha$ from a wide variety of metals with $T-$linear resistivity, based on the Drude model, are consistent with Planckian scattering \cite{corson2000nodal,bruin2013similarity,legros2019universal}. The Planckian limit even applies to conventional metals like gold and copper, where $T-$linear resistivity at high temperatures is caused by electron-phonon scattering. Phonons, however, cannot explain $T-$linear resistivity in the $T\rightarrow0$ limit, suggesting that the Planckian limit is independent of microscopic origin. Estimates based on the Drude model provide no information about how the scattering rate varies in momentum-space. ARPES, on the other hand, provides the momentum dependence \cite{kaminski2005momentum} but only for the single-particle scattering rate, not the transport scattering rate that determines the resistivity. What has been missing is a full momentum-space description of the transport scattering rate.

\begin{figure}[!ht]
\centering
\includegraphics[width=.7\columnwidth]{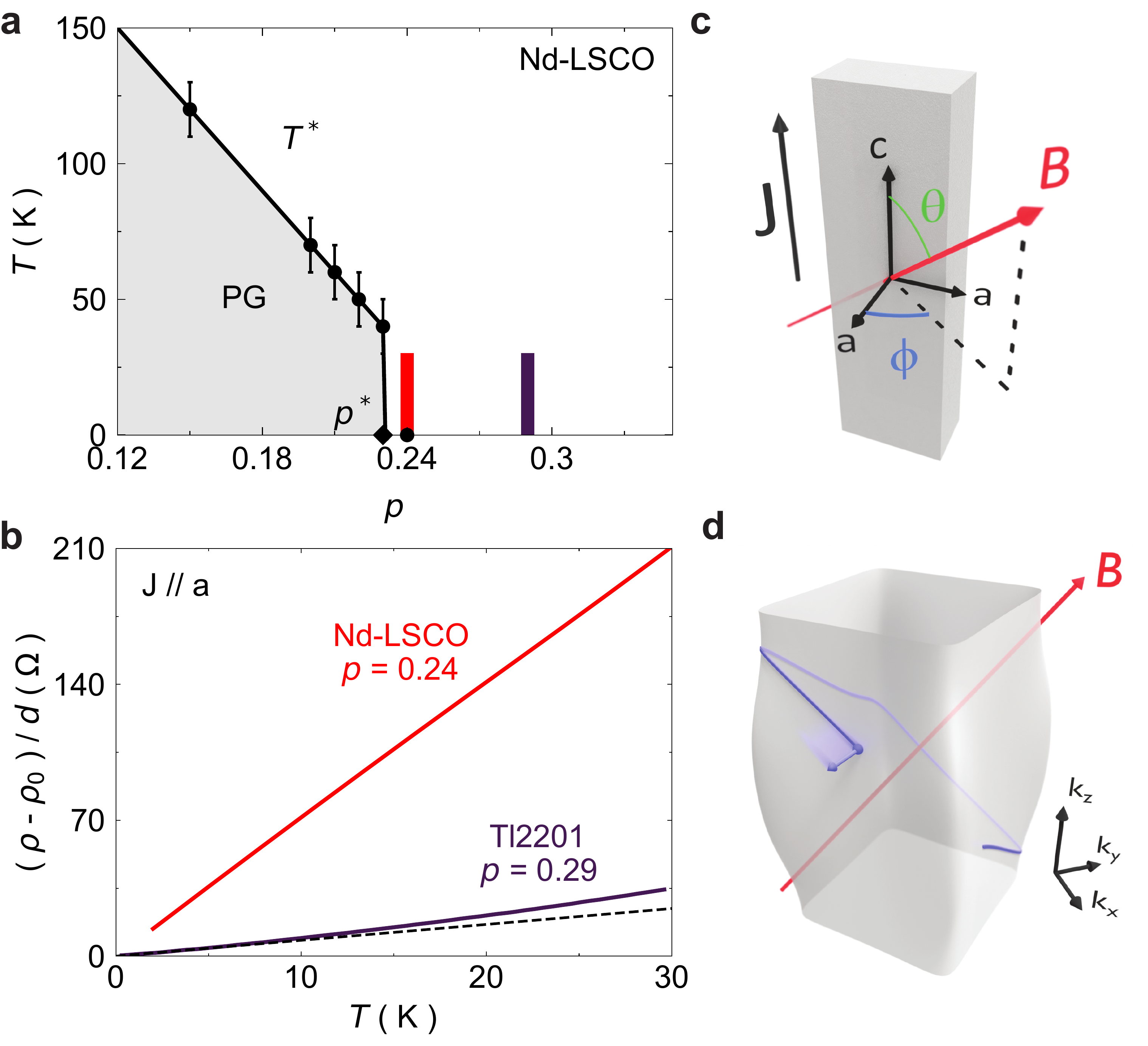}%
\caption{
\textbf{$\bm{T}$-linear resistivity and the angle dependent magnetoresistance technique}.
(\textbf{a})~Temperature-doping phase diagram of the hole-doped cuprate Nd-LSCO.
The pseudogap phase, which onsets below a critical doping of \ps = 0.24 in Nd-LSCO, is highlighted in grey (the onset temperature \Ts of the pseudogap phase is taken from resistivity \cite{daou2009linear,collignon2017fermi}). The superconducting dome is not represented as it can be suppressed with a magnetic field $B \parallel c \ge 20$~T. The red stripe indicates Nd-LSCO at $p=0.24$ measured in the ADMR experiment, the violet stripe represents Tl2201 at $p=0.29$.
(\textbf{b})~In-plane resistivity per copper-oxide plane as a function of temperature for Nd-LSCO at $p=0.24$ at $B=35$~T \cite{daou2009linear} (red) and Tl2201 at $p=0.29$ at $B=13$~T \cite{proust_tl2201} (violet). For both sets of data, the elastic part of the resistivity $\rho_0$ has been subtracted from the total resistivity $\rho$ ($\rho_{\rm xx}$) and divided by the distance $d$ between the CuO$_2$ planes. The black dashed line is the $T$-linear component of the resistivity of Tl2201 $p=0.29$.
(\textbf{c})~Geometry of the ADMR measurement. The sample is represented in gray. The black arrow on the left identifies the direction of the electric current, $\bm{J}$, along the $c$-axis.
The angles $\phi$ and $\theta$ describe the direction of the magnetic field $\bm{B}$ with respect to the crystallographic $a$- and $c$-axis.
(\textbf{d})~The 3D Fermi surface of Nd-LSCO at $p=0.24$ obtained from the ADMR. A single cyclotron orbit, perpendicular to the magnetic field $\bm{B}$, is drawn in blue, with the Fermi velocity indicated with the small blue arrow at a time $t$.
}
\label{fig:phase_diagram}%
\end{figure}

\section*{Technique}

To measure the transport scattering rate in a metal with $T-$linear resistivity, we turn to the \highTc cuprate \NdLSCO (Nd-LSCO) at a hole doping of $p=0.24$. Strange metals are often found in proximity to quantum critical points, and the pseudogap critical point in Nd-LSCO terminates at a hole doping of $\ps=0.23$ as determined by both transport \cite{collignon2017fermi} and ARPES \cite{matt2015electron} measurements (see \autoref{fig:phase_diagram}a). At $p=0.24$, Nd-LSCO shows perfectly \Tlinear resistivity \cite{daou2009linear,collignon2017fermi} down to the lowest measured temperatures once superconductivity is suppressed by a magnetic field (see \autoref{fig:phase_diagram}b).

The technique we use to access the quasiparticle scattering rate is angle-dependent magnetoresistance (ADMR), which measures variations in the $c$-axis resistivity ($\rho_{\rm zz}$) as the sample is rotated to different azimuthal ($\phi$) and polar ($\theta$) angles with respect to an external magnetic field $\bm{B}$(\autoref{fig:phase_diagram}c). The intuitive way of understanding ADMR is to consider that resistivity depends only on the lifetimes and velocities of quasiparticles at the Fermi surface. The application of a magnetic field alters quasiparticle velocities through the Lorentz force, producing variations in the $c$-axis resistivity that depend sensitively on the direction of the magnetic field, hence angle-dependent magnetoresistance. We compare the measured ADMR to calculations made using Chambers' exact solution to the Boltzmann transport equations in a magnetic field \cite{Chambers1952kinetic} and adjust the Fermi surface geometry and the momentum-dependence of the quasiparticle scattering rate in our model until the calculations match the experimental data. This procedure does not assume the presence of a Fermi liquid: Boltzmann transport has been shown to be valid even in cases where Fermi liquid quasiparticles are not present \cite{prange1964transport,abrahams2003hall}.

\section*{Results}

\begin{figure}[!ht]
\centering
\includegraphics[width=.95\columnwidth]{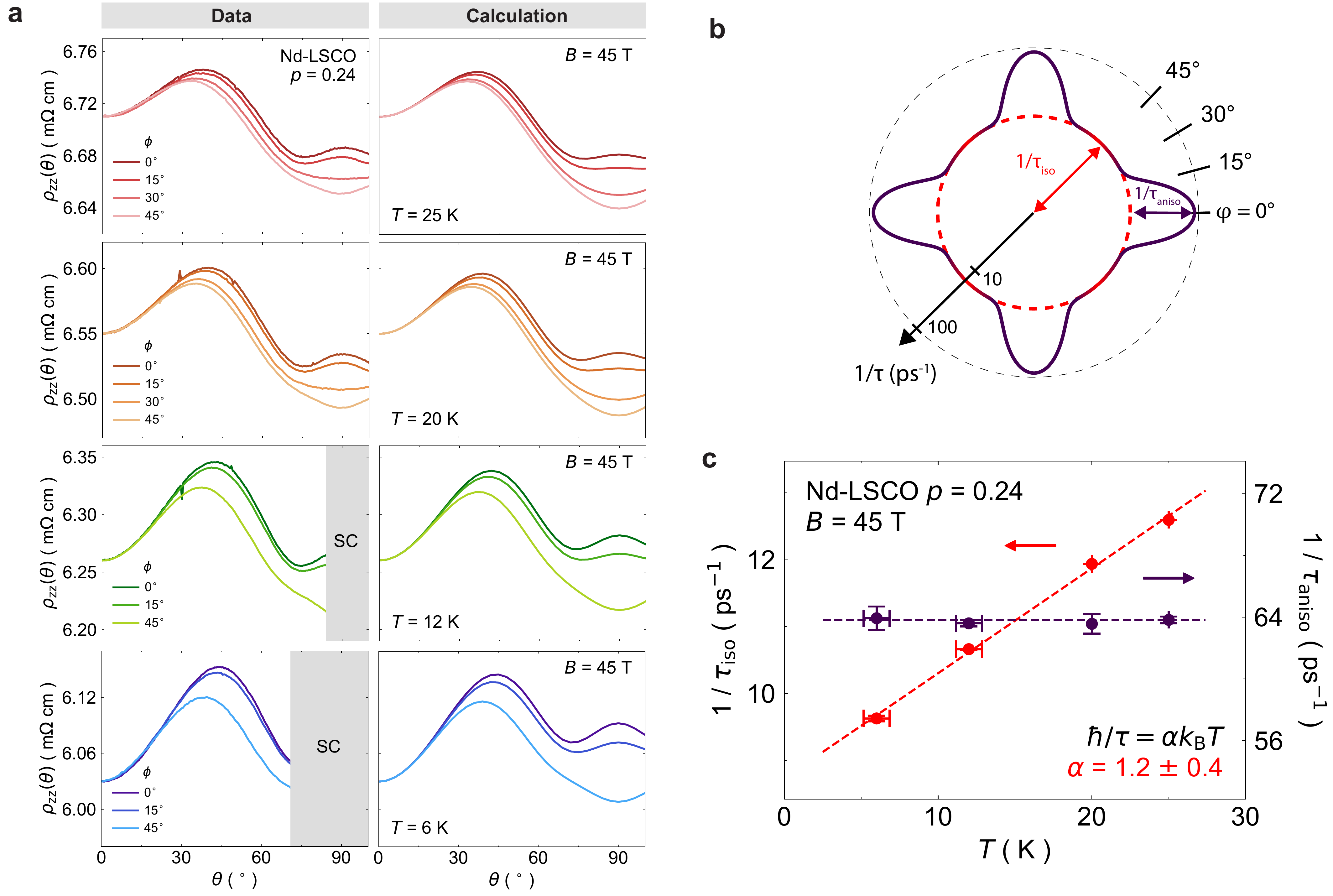}%
\caption{
\textbf{ADMR and quasiparticle scattering rate of Nd-LSCO at $\bm{p = 0.24}$.}
(\textbf{a}) Left panels: The ADMR of Nd-LSCO at $p = 0.24$ as a function of $\theta$ for four different temperatures, $T=25$, $20$, $12$, and $6$~K, and at $B=45$~T. The grey area near $\theta = \degr{90}$ for $T=6$~K and $12$~K indicates the region where the sample becomes superconducting. Right panels: Simulations obtained from the Chambers formula using the tight-binding parameters of \autoref{tab:fit_result_0p24_bandstructure} and the scattering rate model of \autoref{eq:scattering_cosine}.
(\textbf{b}) Log-scale polar plot of the scattering rate at $T=25$~K. Note the large scattering rate near the anti-nodes where the Fermi surface passes close to the van Hove point. The isotropic part of the scattering rate, $1/\tau_{\rm iso}$, is shown as a dashed red line. The anisotropic part, $1/\tau_{\rm aniso}$ is shown in violet. The total scattering rate, $1/\tau_{\rm aniso}+1/\tau_{\rm iso}$ is the entire solid line, shaded red or violet depending on whether it is dominated by $1/\tau_{\rm aniso}$ or $1/\tau_{\rm iso}$, respectively.
(\textbf{c}) Temperature dependence of the two components of the scattering rate. A linear fit to $1/\tau_{\rm iso}$ using $1/\tau = A + \alpha k_{\rm B} T/\hbar$, yields $\alpha=1.2\pm 0.4$, a value consistent with the Planckian limit ($\alpha\approx 1$). The error bar on $\alpha$ accounts for the uncertainty in the fit as well as a $\pm 10~$\% uncertainty in the distance between the electrical contacts on the ADMR sample. By contrast, $1/\tau_{\rm aniso}$ is seen to be temperature independent, showing that it comes entirely from elastic scattering off defects and impurities.
}
\label{fig:admr_scattering}%
\end{figure}

The left panels of \autoref{fig:admr_scattering}a show the ADMR of Nd-LSCO at $p = 0.24$ for $T=6$, $12$, $20$ and $25$~K. These measurements were performed at the National High Magnetic Field Lab using a single-axis rotator to vary the polar angle $\theta$ in a fixed field of 45~T (see \autoref{fig:phase_diagram}c for the experimental geometry). We determine the Fermi surface geometry and the quasiparticle scattering rate by fitting the data simultaneously at all temperatures to a one-band tight-binding model that is commonly used for LSCO-based cuprates (see methods.) We optimize the tight-binding and scattering-rate parameters using a genetic algorithm, taking initial parameter estimates from previous ARPES measurements \cite{matt2015electron,horio2018three}. We set the overall energy scale of the model to be $t = 160 \pm 30 $ meV based on the measured specific heat \cite{michon2019thermodynamic} (see methods). Note that, below 10 K, the specific heat of Nd-LSCO at $p = 0.24$ increases as $\log\left(1/T\right)$ as $T\rightarrow 0$. The resistivity, however, remains linear to low temperature, suggesting that either this correction renormalizes the scattering rate and the bandwidth equally and thus cancels, or the $\log\left(1/T\right)$ factor is not associated with the conduction electrons. As our measurements cannot distinguish between these two scenarios, we omit the $\log\left(1/T\right)$ correction (which would reduce the bandwidth by $\sim 20\%$ at 6 K, see methods).

The simulated ADMR curves produced by these fits are displayed in \autoref{fig:admr_scattering}a (right panels). Key features reproduced by the fit include the position of the maximum near $\theta = \degr{40}$, the onset of $\phi$ dependence beyond $\theta = \degr{40}$, the $\phi$-dependent peak/dip near $\theta = \degr{90}$, and the absolute value of $\rho_{zz}$. The Fermi surface produced by this fit, shown in \autoref{fig:phase_diagram}d, agrees with ARPES measurements \cite{matt2015electron,horio2018three}. The best-fit tight-binding parameters are the same as those determined by ARPES to within our uncertainty (see \autoref{tab:fit_result_0p24_bandstructure}), demonstrating remarkable consistency between the two techniques. 

We now consider the scattering rate obtained from the fit. We separate the scattering rate in our model into two components---one isotropic and one anisotropic: $1 / \tau(\bm{k}) = 1/\tau_{\rm iso}+1/\tau_{\rm aniso}(\bm{k})$. We find that the ADMR is best described by a highly anisotropic scattering rate that is largest near the ``anti-nodal'' ($\phi=\degr{0}$, \degr{90}, \degr{180} and \degr{270}) regions of the Brillouin zone and smallest near the ``nodal'' ($\phi=\degr{45}$, \degr{135}, \degr{225} and \degr{315}) regions (\autoref{fig:admr_scattering}b). In \autoref{fig:scattering_models_0p24}, we show that three different phenomenological models of $1/\tau(\bm{k})$ all converge to the same shape as a function of $\phi$, indicating that our fit is independent of the specific function chosen (see \autoref{fig:admr_scattering}b.)

We extract the scattering rate at each temperature by fitting the full $\theta$- and $\phi$-dependent $\rho_{zz}$, and we \textit{a priori} assume no particular temperature dependence---the scattering-rate parameters are determined independently at each temperature, while the FS geometry parameters are held constant). We extract the temperature dependence of both the isotropic and anisotropic components of the quasiparticle scattering rate from these fits, shown in \autoref{fig:admr_scattering}c. Remarkably, we find that the anisotropic scattering rate is temperature independent, while the isotropic scattering rate is linear in temperature.

\begin{figure}[h!]%
\centering
\includegraphics[width=.85\columnwidth]{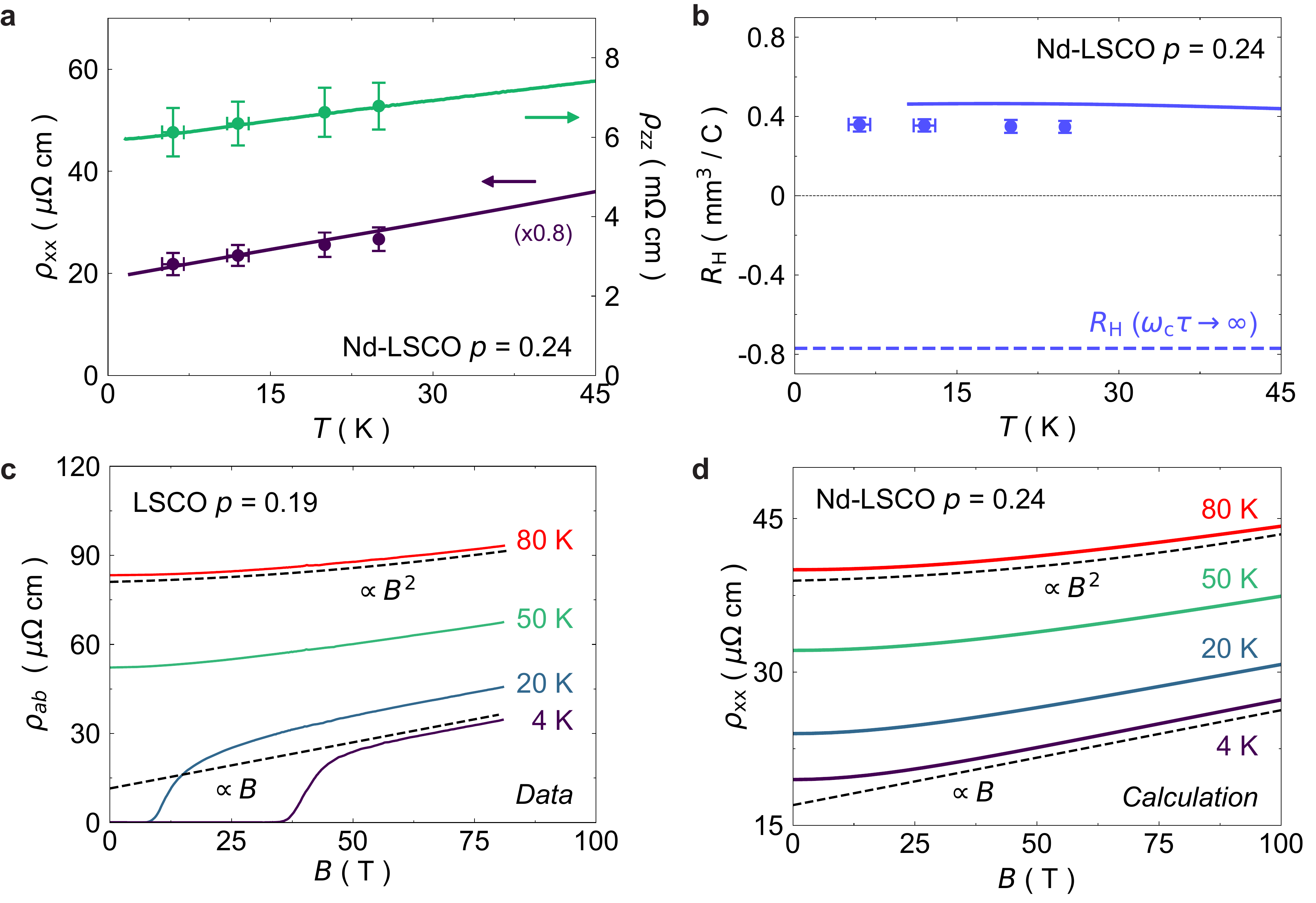}%
\caption{\textbf{Transport coefficients of Nd-LSCO at~$\bm{p=0.24}$.}
(\textbf{a}, \textbf{b})~$\rho_{\rm xx}$ and $\rho_{\rm zz}$ at $B=33$~T and $R_{\rm H}$ at $B=15$~T, respectively. Solid lines represent the data measured on Nd-LSCO at $p=0.24$ ($\rho_{\rm xx}$, $\rho_{\rm zz}$ and $R_{\rm H}$ \cite{daou2009linear}). Circles represent the values calculated using the scattering rates plotted in \autoref{fig:admr_scattering}c. The $\rho_{\rm xx}$ data are taken on a different sample \cite{daou2009linear} to that used in the AMDR measurements and from which the scattering rates are extracted. While systematic errors on geometric factors are expected from sample to sample, it is seen that a constant factor of $0.8$ on the data is sufficient to give excellent agreement between calculation and data. The dashed blue line in panel (b) represents the high-field ($\omega_{\rm c}\tau\rightarrow \infty$) limit for the Fermi surface of Nd-LSCO at $p=0.24$. The difference between this limit and the data comes from the small value of $\omega_c \tau = 0.024$ at $T=25$~K and $B=45$~T and the fact that the conductivity is highest in the nodal directions where the Fermi surface has a hole-like curvature (\autoref{fig:ndlsco_vs_tl2201}a).
(\textbf{c}, \textbf{d})~In-plane resistivity as a function of magnetic field, with data from LSCO at $p=0.19$ (just above its own pseudogap critical point at \ps=0.18) \cite{giraldo2018scale} on the left and calculations using the scattering rate values obtained from the ADMR data on Nd-LSCO at $p=0.24$ (extrapolated linearly to 100 K) on the right. In our calculations we find $B$-linear magnetoresistance at low temperature (dashed line) that becomes $B^2$ at high temperature (dashed line), as observed in LSCO $p = 0.19$.
}%
\label{fig:admr_transport}%
\end{figure}

To check the validity of these scattering rates, we use our fit parameters and Bolzmann transport to calculate the temperature dependence of $\rho_{xx}$ and the Hall coefficient $R_{\rm H} \equiv \rho_{xy}/B$. As shown in \autoref{fig:admr_transport}, we reproduce the temperature dependence of all three transport coefficients. While the Fermi surface at $p=0.24$ is electron-like (i.e. it is centered on the $\Gamma$ point in the first Brillouin zone), both the measured and calculated $R_{\rm H}$ are hole-liked due to the Fermi surface curvature \cite{narduzzo2008violation} (see \autoref{fig:admr_transport}b). An anisotropic scattering rate, highly enhanced near the anti-nodal regions of the Fermi surface (\autoref{fig:admr_scattering}b and \ref{fig:scattering_models_0p24}), is therefore not only required to correctly model the ADMR, but also required to obtain the correct sign and magnitude of the Hall coefficient. To ensure that our fits are not fine-tuned for $B = 45$ T, we fit a second data set taken at $B = 35$~T (\autoref{fig:admr_35}). We fix the tight-binding parameters to those obtained from the 45 T fits, and we find that the same scattering-rate parameters emerge at 35 T, demonstrating the consistency of the model.

\section*{Discussion}

\begin{figure}[h!]
\centering
\includegraphics[width=.9\columnwidth]{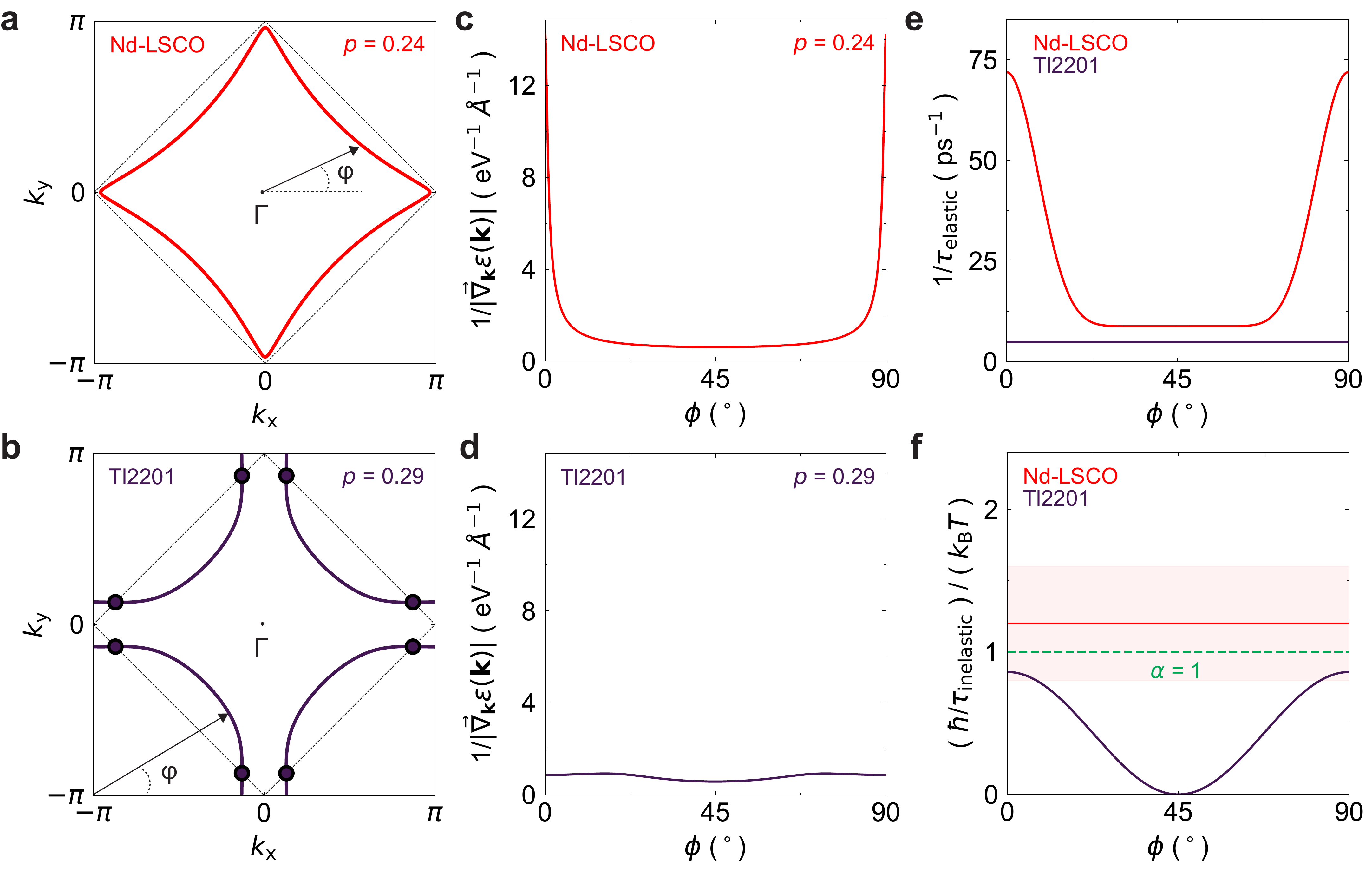}%
\caption{\textbf{Comparison of two overdoped cuprates – Nd-LSCO and Tl2201} | (\textbf{a}, \textbf{b}) Fermi surfaces at $k_{\rm z} = \pi / c$. In Nd-LSCO at $p = 0.24$ ((a), red), the Fermi surface is electron-like and contained inside the antiferromagnetic zone boundary (black dotted lines). In Tl2201, with $T_{\rm c} = 15$~K ((b), violet), the hole-like Fermi surface crosses the antiferromagnetic zone boundary at so-called “hot spots” (violet points). (\textbf{c}, \textbf{d}) Density of states (DoS) $1/|\vec{\nabla}_{\bm{k}}\epsilon(\bm{k})|$ as a function of the azimuthal angle $\phi$, at $k_{\rm z} = \pi / c$. In Nd-LSCO (c), the DoS is large at the antinodes due to proximity to the van Hove singularity. By contrast, in Tl2201 (d), the DoS is nearly isotropic. (\textbf{e}) Elastic part of the scattering rate vs azimuthal angle $\phi$. In Nd-LSCO (red), the elastic scattering rate tracks the strong angle dependence of the DoS. By contrast, the elastic scattering rate in Tl2201 (violet; from \cite{abdel2006anisotropic}) is isotropic, in accordance with the relatively isotropic DoS. (\textbf{f}) Inelastic part of the scattering rate, multiplied by $\hbar/(k_{\rm B} T)$, vs azimuthal angle $\phi$. The inelastic scattering rate in Nd-LSCO is isotropic and consistent with “Planckian dissipation” in the sense that $\hbar/\tau = \alpha k_{\rm B} T$ with $\alpha$ of order 1 (the uncertainty in $\alpha$ is indicated by the red shading.) The inelastic $T$-linear scattering rate of Tl2201 is strongly anisotropic, going from zero at $\phi=\degr{45}$ (nodal region) to a near-Planckian magnitude at $\phi=\degr{0}$ (anti-nodal region, near the hot spots). Note that in Tl2201 there is also an isotropic $T^2$ part to the inelastic scattering rate, in addition to the anisotropic $T$-linear part shown here \cite{abdel2006anisotropic}. This results in a resistivity that varies as $a T + b T^2$ (\autoref{fig:phase_diagram}b) \cite{proust_tl2201}.}%
\label{fig:ndlsco_vs_tl2201}%
\end{figure}

We have measured the momentum dependence of the scattering rate responsible for the \Tlinear resistivity of Nd-LSCO at $p = 0.24$. We can write the total scattering rate as a sum of an elastic (temperature-independent) component plus an inelastic (temperature-dependent) component \footnote{We use the working-definitions of ``elastic scattering'' to mean temperature-independent scattering and ``inelastic scattering'' to mean temperature-dependent scattering. There are exceptions to these definitions but they hold under most cases, particularly in the low-temperature limit.}:
\begin{equation}
1/\tau\left(\phi, T\right) = 1/\tau_{\rm elastic} + 1/\tau_{\rm inelastic}(T).
\label{eq:tausep}
\end{equation}
We find that $1/\tau_{\rm elastic} =  1/\tau_{\rm aniso}\left(\phi\right) + 1/\tau_{\rm iso}\left(T=0\right) $, i.e. the elastic scattering contains all of the anisotropic scattering, plus the $T = 0$ offset from the isotropic scattering. The elastic term is, by definition, temperature-independent, and its angle dependence resembles the strongly $\phi$-dependent density of states at $p=0.24$ (see \autoref{fig:ndlsco_vs_tl2201}c and e). It was previously suggested that similar anisotropy in the \textit{single}-particle scattering rate (i.e. the scattering rate measured by ARPES) may arise due to the proximity of the anti-nodal Fermi surface to the van Hove singularity \cite{abrahams2000angle}. Our data suggest that a similar anisotropy extends to the \textit{two}-particle transport scattering rate. Indeed, the momentum dependence of the elastic scattering rate we measure is reminiscent of the elastic scattering rate extracted by ARPES in LSCO at $p=0.23$ \cite{chang2013anisotropic}, as shown in Supplementary Information Figure S2.

We find that the inelastic term in \autoref{eq:tausep} has a pure \Tlinear dependence whose strength is consistent with Planckian dissipation, i.e. $1/\tau_{\rm inelastic}(T) = \alpha \frac{k_B T}{\hbar}$, with $\alpha$ close to 1 (see \autoref{fig:ndlsco_vs_tl2201}f.) This unambiguously demonstrates that \Tlinear resistivity is caused by a \Tlinear scattering rate and not, for example, by a $T$-dependent carrier density \cite{pelc2020resistivity}. Remarkably, we discover that this Planckian scattering is isotropic---the same for all directions of electron motion. Isotropic, \Tlinear scattering has been hypothesized in the context of a marginal Fermi liquid description of the normal state of cuprates \cite{varma1989phenomenology}. The marginal Fermi liquid also hypothesizes an $\omega$-linear scattering rate, and this was observed by ARPES in LSCO \cite{kaminski2005momentum}. The absence of momentum-space structure to the scattering rate implies that the microscopic mechanism of \Tlinear resistivity is length-scale invariant, i.e. it does not depend on scattering from a particular wavevector, such as the fluctuations of a finite-$q$ order parameter. The fact that the inelastic scattering rate appears to reach a limit dictated by Planck's constant suggests that a fundamental quantum principle is at play, akin to that involved in the maximal rate of entropy production at a black hole event horizon \cite{kovtun2005viscosity}. As was found in previous studies \cite{bruin2013similarity}, the Planckian limit constrains only the temperature dependent part of the scattering rate.

Detailed knowledge of the Fermi surface and the scattering rate allows us to examine other transport properties in more quantitative detail than was previously possible. In \autoref{fig:admr_transport}d, we plot the calculated isotherms of $\rho_{\rm xx}$ versus $B$ up to 100~T. We see that a strong $B$-linear component is present at low $T$ above a threshold field, whereas a quadratic $B^2$ dependence dominates at high~$T$ and low~$B$---strikingly similar to what was recently measured in LSCO \cite{giraldo2018scale} (\autoref{fig:admr_transport}c). This $B-$linear magnetoresistance occurs naturally at intermediate fields between the low-field $B^2$ regime and the field-independent regime that occurs once $\omega_c \tau >>1$ \cite{peierls1931theorie} (see SI for more details). When $v_F$ or $\tau$ are highly anisotropic, as is the case for Nd-LSCO at $p = 0.24$ and LSCO at $p = 0.19$, the high-field regime is pushed up to extremely high fields, resulting in a broad region of $B-$linear magnetoresistance. This mechanism may explain $B-$linear magnetoresistance without any need for a $B$-dependent scattering rate. This is further supported by our fits to a second data set taken at $B = 35$ T, which yield the same scattering rates we find at 45 T (\autoref{fig:admr_35}.) It remains to be seen whether this mechanism can explain $B-$linear magnetoresistance more generally, e.g. as found in iron pnictide superconductors \cite{hayes2016scaling}, where the Fermi surface and scattering rate are unlikely to be as anisotropic as they are in Nd-LSCO.


In the context of our discovery that the inelastic scattering rate at \ps is both Planckian and isotropic, it is interesting to consider how this scattering rate evolves into the overdoped regime. Far above \ps, for example in LSCO at $p$ = 0.33 \cite{nakamae2003electronic}, the resistivity is $T^2$, as expected for a Fermi liquid. As the doping is lowered toward \ps, the $T^2$ component of the resistivity shrinks while a \Tlinear contribution grows \cite{cooper2009anomalous}. Prior ADMR studies on overdoped Tl$_2$Ba$_2$CuO$_{\rm 6+\delta}$ (Tl2201), at $p = 0.29$ (\Tc$ = 15$ K) \cite{abdel2006anisotropic}, have found coexistence between an isotropic $T^2$ scattering rate and an anisotropic \Tlinear scattering rate (see \autoref{fig:ndlsco_vs_tl2201}d, e and f), agreeing with the temperature dependence of the resistivity in Tl2201 (\autoref{fig:ndlsco_vs_tl2201}b). While ADMR has not been performed in a single material at both \ps and in the highly overdoped regime, a useful comparison can be made between Tl2201 and Nd-LSCO.

First we compare the elastic scattering rate, which is isotropic in Tl2201 versus strongly anisotropic in Nd-LSCO (\autoref{fig:ndlsco_vs_tl2201}e). We attribute this to a difference in the density of states: nearly isotropic in Tl2201 (\autoref{fig:ndlsco_vs_tl2201}b and d) versus strongly anisotropic in Nd-LSCO due to the proximity of its FS to the van hove point (\autoref{fig:ndlsco_vs_tl2201}a and c). The second, more interesting difference between the two materials is in the inelastic scattering rate. In Nd-LSCO, the inelastic scattering rate is entirely \Tlinear and has the full Planckian magnitude for all $k$ directions (\autoref{fig:ndlsco_vs_tl2201}f). By contrast, the inelastic scattering in Tl2201 is only in part \Tlinear, and the magnitude of the \Tlinear component is only Planckian along the anti-nodal directions (\autoref{fig:ndlsco_vs_tl2201}f). As a result, the resistivity of Tl2201 is not \Tlinear, varying as $aT + bT^2$, with a \Tlinear component one order of magnitude smaller than in Nd-LSCO (\autoref{fig:phase_diagram}b). This comparison suggests that for a metal to display a pure \Tlinear resistivity, its scattering rate must grow to reach the Planckian limit for all directions. This could explain why a pure \Tlinear resistivity can be found in metals with vastly different Fermi surfaces, \textit{e.g.} quasi-1D single-band organic metals like the Bechgaard salts \cite{doiron2009correlation} and 3D multi-band $f$-electron metals like CeCu$_{\rm 5.9}$Au$_{\rm 0.1}$ \cite{lohneysen1994}. ADMR studies on Tl2201 at lower doping would prove invaluable: we predict that the \Tlinear component of the scattering rate will grow in the nodal directions to become isotropic at \ps, while the $T^2$ component will decrease and then vanish at that same doping.

\def\bibsection{\section*{\refname}}

\bibliographystyle{unsrtnat}
\bibliography{ms-admr_ndlsco_0p24_v4}

\section*{Acknowledgments}
The authors acknowledge helpful discussions with James Analytis, Debanjan Chowdhury, Nicholas Doiron-Leyraud, Nigel Hussey, Mark Kartsovnik, Steve Kivelson, Dung-Hai Lee, Sylvia Lewin, Akash Maharaj, Andr$\acute{\rm e}$-Marie Tremblay, Kimberly Modic, Chaitanya Murthy, Seth Musser, Cyril Proust, Arkady Shekhter, Senthil Todadri, Chandra Varma.
A portion of this work was performed at the National High Magnetic Field Laboratory, which is supported by the National Science Foundation Cooperative Agreement No. DMR-1644779 and the State of Florida.
PAG acknowledges that this project is supported by the European Research Council (ERC) under the European Union’s Horizon 2020 research and innovation program (Grant Agreement No. 681260).
J.-S.Z. was supported by an NSF grant (MRSEC DMR-1720595).
L.T. acknowledges support from the Canadian Institute for Advanced Research (CIFAR) as a Fellow and funding from the Natural Sciences and Engineering Research Council of Canada (NSERC; PIN: 123817), the Fonds de recherche du Qu\'ebec - Nature et Technologies (FRQNT), the Canada Foundation for Innovation (CFI), and a Canada Research Chair. This research was undertaken thanks in part to funding from the Canada First Research Excellence Fund. Part of this work was funded by the Gordon and Betty Moore Foundation’s EPiQS Initiative (Grant GBMF5306 to L.T.)
B.J.R. and Y.F. acknowledge funding from the National Science Foundation under grant no. DMR-1752784.

\section*{Author Contributions}

A.L., P.G., L.T., and B.J.R. conceived the experiment.
J.-S.Z. grew the sample.
A.L., F.L., A.A., C.C. and M.D. performed the sample preparation and characterization.
G.G., Y.F., A.L., D.G., P.G., and B.J.R. performed the ADMR measurements at the National High Magnetic Field Laboratory in Tallahassee.
G.G., Y.F., S.V., M.J.L., and B.J.R. performed the data analysis and simulations.
G.G., L.T., and B.J.R. wrote the manuscript with input from all other co-authors.
L.T. and B.J.R. supervised the project.

\section*{Competing Interests}
The authors declare no competing interests.

\section*{Methods}

\subsection*{Samples and Transport Measurements}
\renewcommand{\figurename}{Extended Data Fig.}
\newcounter{extended}
\setcounter{extended}{0}
\renewcommand\theextended{\arabic{extended}}
\renewcommand\thefigure{\arabic{extended}}

Single crystal La$_{\rm 2-y-x}$Nd$_{\rm y}$Sr$_{\rm x}$CuO$_{\rm 4}$ (Nd-LSCO) was grown at the University of Texas at Austin using a traveling-float-zone technique, with a Nd content $y = 0.4$ and nominal Sr concentration $x = 0.25$. The hole concentration is $p = 0.24 \pm 0.005$ (for more details, see ref. \cite{collignon2017fermi}). The value of $T_{\rm c}$, defined as the point of zero resistance, is $T_{\rm c} = 11$~K. The pseudogap critical point in Nd-LSCO is at $p^* = 0.23$ (ref. \cite{collignon2017fermi}).

\begin{figure}[h!]%
\centering
\includegraphics[width=1\columnwidth]{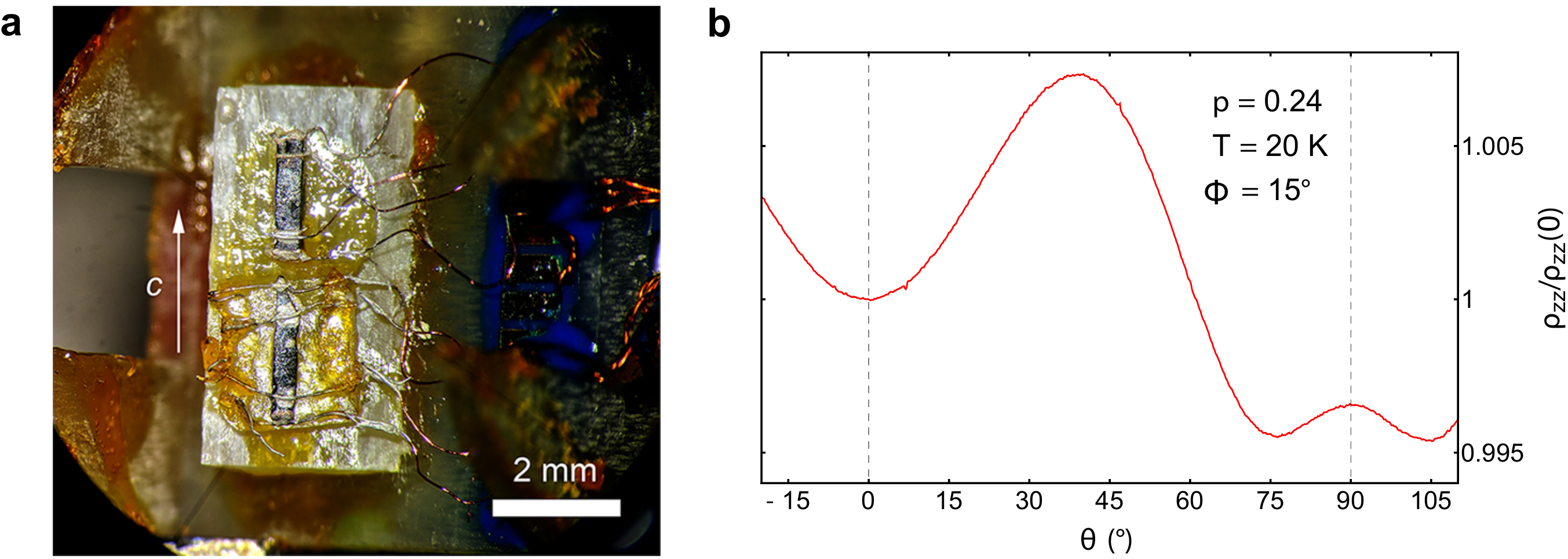}
\refstepcounter{extended}\label{fig:sampleMount}
\caption{\textbf{ADMR experimental set up.} (\textbf{a}) A photograph of the sample on the rotator. The two samples here are mounted on a G-10 wedge to provide a $\phi$ angle of \degr{30}. Additional wedges provided angles of $\phi = \degr{15}$ and $\degr{45}$; (\textbf{b})~ADMR as a function of $\theta$ angle from \degr{-15} to \degr{110} and $\phi=0$ at $T=20$~K for Nd-LSCO $p=0.24$, showing the symmetry of the data about these two angles.}
\end{figure}

Resistivity measurements were performed in the 45~T hybrid magnet at the National High Magnetic Field Lab in Tallahassee, USA. The sample resistance was measured with a standard 4-point contact geometry using a Stanford Research 830 Lock-In Amplifier. The samples were driven with $I_{RMS} = 1$~mA from a Keithley 6221 Current Source. Temperature was stabilized to within $\pm 1$ mK around the target temperature at each angle. Uncertainty of the absolute temperature due to thermometer magnetoresistance is approximately $\pm 1$~K at both $T=6$~K and $T=12$~K (horizontal error bars in \autoref{fig:admr_scattering}c and \ref{fig:admr_transport}a, b), but negligible at $T=20$~K and above.

At $p = 0.24$ the upper critical field of Nd-LSCO is 10~T for $\bm{B} \parallel c$ \cite{michon2019thermodynamic}. By applying a magnetic field of $B=45$~T at both $T = 25$~K and $T = 20$~K the sample remains in the normal state while rotating the field from $\bm{B} \parallel c$ to $\bm{B} \parallel a$. At $T=12$~K and $T=6$~K the $p=0.24$ sample is in the normal state when $\bm{B} \parallel c$, but superconductivity onsets when the field is rotated toward $\bm{B} \parallel a$, as shown in \autoref{fig:admr_scattering}a.

The polar angle $\theta$ between the crystalline $c$-axis and the magnetic field was changed \textit{in situ} continuously from $\approx \degr{-15}$ to $\approx \degr{110}$  using a single-axis rotator (see \autoref{fig:sampleMount}a). A voltage proportional to the angle was recorded with each angle sweep. The angle $\theta$ was calibrated by finding symmetric points in the resistivity and scaling the measured voltage such that the symmetric points lie at $\theta = \degr{0}$ and $\degr{90}$ (see \autoref{fig:sampleMount}b). This procedure resulted in an uncertainty in $\theta$ of $\pm \degr{0.5}$. The azimuthal angle $\phi$ was changed by placing the sample on top of G-10 wedges machined at different angles: \degr{15}, \degr{30} and \degr{45}. An illustration of the sample mounted on the rotator stage, with a G-10 wedge to set the azimuthal angle to be \degr{30}, is shown in \autoref{fig:sampleMount}. The samples and wedges were aligned under a microscope by eye to an accuracy in $\phi$ of $\pm \degr{2}$.

\subsection*{Fitting method}
\textbf{Genetic algorithm}. Computing the conductivity as described above involves free parameters (e.g. $t'$, $t''$, $t_z$, $\mu$, $\tau_{\rm iso}$, $\tau_{\rm aniso}$, $\nu$) which can be written as a vector $\bm x$. The optimal $\bm x$, which we refer to as $\bm {x^*}$, minimizes the chi-square ($\chi^2$) statistic between the resistivity from the model $\rho_{\rm zz}^{model}(\bm x, \theta, \phi)$ and the measured resistivity $\rho^{\text{data}}_{\rm zz}(\theta, \phi)$ at all magnetic field orientations $(\theta,\phi)$:

\begin{align}
\chi^2(\bm x)
&=
\sum_{(\theta,\phi)}
\big(
\rho_{\rm zz}^{model}(\bm x, \theta, \phi)
- \rho^{\text{data}}_{\rm zz}(\theta, \phi)
\big )^2,
\end{align}
We thus seek $\bm x^*$ such that:
\begin{align}
\bm x^{*} = \arg\min_{\bm x} \chi^2(\bm x).
\end{align}

Using the Chambers formula to fit the ADMR measurements can be tricky for standard optimization algorithms such as gradient based methods. They are either slow to converge, highly sensitive to the initial conditions, or most annoyingly they tend to get stuck in local minima of the $\chi(\bm x)$ landscape. That is the reason why we turned to a genetic algorithm (or ``differential evolution'') as a global optimization method which can avoid these issues. The genetic algorithm has become a standard fitting routine in science, it is carefully detailed in the supplementary information of \cite{ramshaw2017broken}. For this study we used the differential evolution algorithm from the Python package \textit{lmfit} \cite{lmfit} and our own C++ implementation. We back checked the efficiency of the genetic algorithm with two other global optimizers, such as AMPGO (Adaptive Memory Programming for Global Optimization) and SHGO (Simplicial Homology Global Optimization) also made available in \textit{lmfit} \cite{lmfit}. The three optimizers all converged to the same results, confirming the robustness of our fit procedure.

\textbf{Convergence criteria}. The $\chi^2$ values of each member of the population are calculated after each generation of optimization. The distribution of all these $\chi^2$ values follows a Gaussian-like distribution. The genetic algorithm stops when the standard deviation of this distribution has reached less than 1\% of the mean value of the distribution.

\textbf{Error bars}. When the fit reaches $\bm{x}^*$ (the best fit values) the error bars are calculated for each parameter by the statistical procedure of calculating the Hessian matrix, which represents the second derivative of the fit quality $\chi^2$ in regard to each parameter. The error bars in \autoref{tab:fit_result_0p24_scattering} are calculated as the square root of the diagonal values of the covariance matrix (inverse of the Hessian matrix) evaluated at $\bm{x}^*$. More details are available on the website of the Python package \textit{lmfit} \cite{lmfit}.

\textbf{Fitting procedure}. To find the tight-binding and scattering rate parameter values that best describe the Fermi surface of Nd-LSCO at $p=0.24$ at all temperatures, we searched the parameter space using the following:
\begin{enumerate}
  \item We fit the ADMR data at the four temperatures (6, 12, 20, 25~K) simultaneously with the genetic algorithm. All temperatures share the same tight-binding parameters during the optimization process, but the scattering rate parameters ($1/\tau_{\rm iso}$, $1/\tau_{\rm aniso}$, $\nu$ for the cosine model \autoref{eq:scattering_cosine}) are unique for each temperature.
  \item The search range of the genetic algorithm for the tight-binding parameters was set at $\pm 30$~\% around the ARPES values provided by Johan Chang through private communications for the data presented in \cite{matt2015electron}: $\mu=-0.93t$, $t'=-0.136t$, $t''=0.068t$, $t_{\rm z}=0.07t$ (this last value comes from \cite{horio2018three} for Eu-LSCO which shows identical atomic structure and electronic properties). Only $t=190$~meV was kept fixed.
  \item The absolute value of $\rho_{\rm zz}$ at each temperature---not just the relative change with angle---was included in the optimization.
\end{enumerate}

\begin{table}[h!]
  \begin{center}
  \begin{adjustbox}{width=1\textwidth}
    \begin{tabular}{|c|c|c|c|c|c|c|}
    \hline
     & $t$ (meV) & $t'/t$ & $t''/t$ & $t_z/t$ & $\mu/t$ & $p$\\
    \hline
    ADMR  & $160 \pm 30$ & $-0.1364\pm0.0005$ & $0.0682\pm0.0005$ & $0.0651\pm0.0005$ & $-0.8243\pm0.0005$ & 0.248\\
    ARPES & 190 & -0.136 & 0.068 & 0.07 &  & 0.28 \\

    \hline
    \end{tabular}
    \end{adjustbox}
  \end{center}
 \caption{\textbf{Tight-binding parameters from the fit to the ADMR~data at $\bm{p}$ = 0.24 .} Best fit tight-binding values for the Nd-LSCO $p=0.24$ ADMR data (using the cosine scattering rate model of \autoref{eq:scattering_cosine}). The hopping parameter $t=160~\pm~30$~meV was fixed by the measured specific heat: see the section ``Determining the energy scale $t$'' for more information. The results are extremely close to ARPES tight-binding values reported in \citet{matt2015electron} and \citet{horio2018three}, reproduced here on the second line. Error bars on the AMDR-derived values were obtained following the procedure described in the above section. The error bar on the value of $t_z$ measured by ARPES is $\pm 0.02t$ (J. Chang and M. Horio, private communication).}
    \label{tab:fit_result_0p24_bandstructure}
\end{table}

\subsection*{Band structure}
We use a three dimensional tight binding model of the Fermi surface that accounts for the body-centered tetragonal crystal structure of Nd-LSCO \cite{horio2018three},
\begin{equation}
  \begin{aligned}
\epsilon(k_x,k_y,k_z)=&-\mu
-2t[\cos(k_xa)+\cos(k_ya)]
\\&-4t'\cos(k_xa)\cos(k_ya)
-2t''[\cos(2k_xa)+\cos(2k_ya)]
\\&-2t_z\cos(k_xa/2)\cos(k_ya/2)\cos(k_zc/2)[\cos(k_xa)-\cos(k_ya)]^2,
  \end{aligned}
    \label{eq:band_structure_0p24}
\end{equation}
where $\mu$ is the chemical potential, $t$, $t'$, and $t''$ are the first, second, and third nearest neighbor hopping parameters, $t_z$ is the inter-layer hopping parameter, $a=3.75$~\AA~is the in-plane lattice constant of Nd-LSCO, and $c/2=6.6$~\AA~is the CuO$_2$ layer spacing. The inter-layer hopping has the form factor $\cos(k_xa/2)\cos(k_ya/2)[\cos(k_xa)-\cos(k_ya)]^2$, which accounts for the offset copper oxide planes between layers of the body-centered tetragonal structure \cite{chakravarty1993interlayer}.

The fit results of the ADMR data are presented in \autoref{fig:admr_scattering}a, \autoref{tab:fit_result_0p24_bandstructure} (for the tight-binding values), and \autoref{tab:fit_result_0p24_scattering} (for the scattering rate values). Although the genetic algorithm was allowed to search over a wide range of parameters, we found that the optimal solution converged towards $t'$, $t''$ and $t_{\rm z}$ values extremely close to the ARPES values, with a 7\% deviation at most for $t_{\rm z}$. Only $\mu$, and therefore the doping $p$, is substantially different from the ARPES value. The higher doping found by ARPES may be due to the difficulty in accounting for the $k_z$ dispersion, or may be due to different doping at the surface. Nevertheless, the shape of the Fermi surface found by fitting the ADMR data (see \autoref{fig:phase_diagram}d) is electron like and qualitatively identical to the one measured by ARPES \cite{matt2015electron}, and the doping we find ($p=0.248$) is very close to the nominal one $p=0.24 \pm 0.005$ \cite{daou2009linear}.

This demonstrates that the Fermi surface is correctly mapped out by the ADMR data. In the figures and the analysis presented in this manuscript, we use the tight-binding values from \autoref{tab:fit_result_0p24_bandstructure}, and for simplicity we refer to them as the ``tight-binding values from ARPES'', as they only differ by the chemical potential value.

\subsection*{Determining the energy scale $t$}

Fitting ADMR to a tight binding model using Boltzmann transport determines the relative variation between the different tight-binding parameters. The overall scale $t$, however, must be determined independently. While ARPES can determine $t$ by fitting the measured dispersion to a tight binding model, ARPES does not necessarily have the sensitivity to determine all band renormalizations at the Fermi energy. As electrical transport is \textit{only} sensitive to renormalizations at the Fermi energy, and not to the overall bandwidth, it is crucial to determine $t$ accurately if one is to quantitatively determine the scattering rate. The experimentally-determined quantity that is most sensitive to band renormalizations near the Fermi energy is the specific heat, which is sensitive to the total density of states. To determine $t$, we calculate the density of states from our tight-binding model and adjust $t$ to match the experimentally-determined electronic specific heat, $C_{\rm el}$.

\begin{figure}[h!]
\centering
\includegraphics[width=0.9\columnwidth]{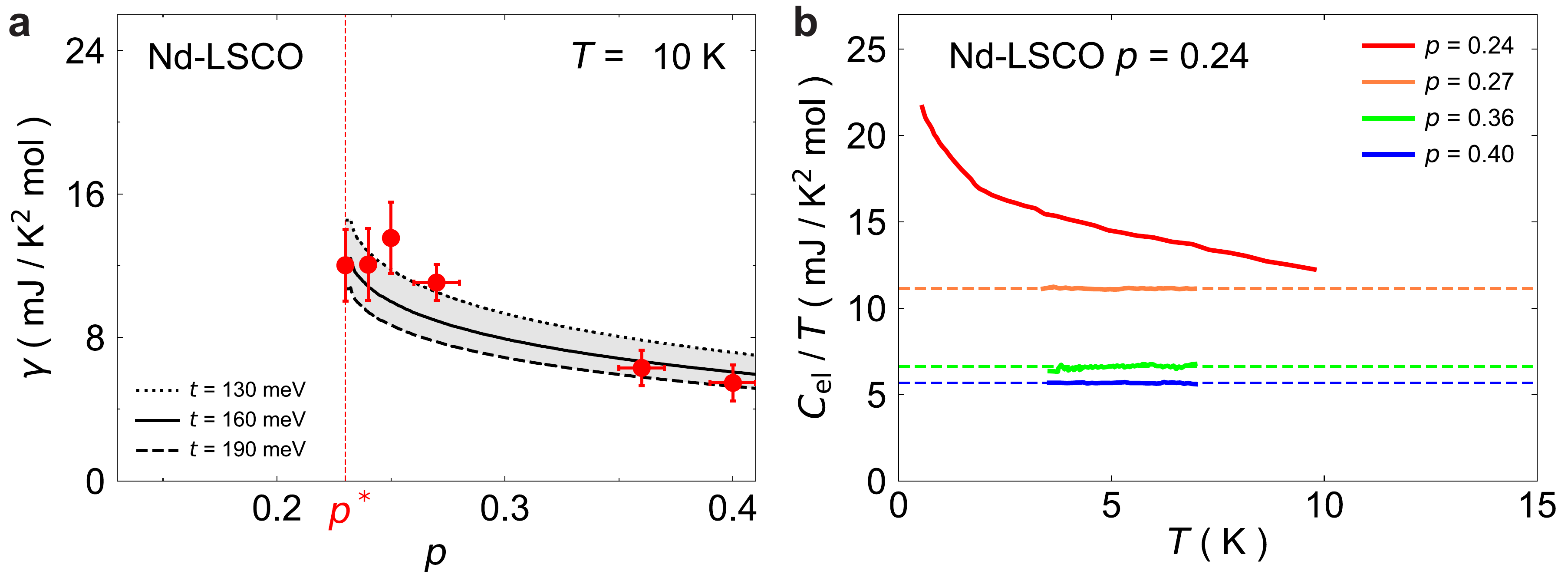}%
\refstepcounter{extended}\label{fig:choice_of_t}
\caption{\textbf{Calculated and measured Sommerfeld coefficients of Nd-LSCO.} (\textbf{a}) The Sommerfeld coefficient $\gamma$ for Nd-LSCO as a function of doping. The measured values (red circles) are obtained from electronic specific heat measurements $C_{\rm el}/T$ \cite{michon2019thermodynamic}. For the calculated $\gamma$ (black solid and dash lines), we use the tight-binding parameters from our ADMR analysis for three different values of $t$, as indicated. The grey band represents the region of consistency between the calculations and the data. (\textbf{b}) Electronic specific heat $C_{\rm el}/T$ as a function of temperature for Nd-LSCO $p=0.24$, $0.27$, $0.36$ and $0.40$ \cite{michon2019thermodynamic}. The data are the solid lines and the dashed lines represent extrapolations.
}
\end{figure}

In \autoref{fig:choice_of_t}a, we compare the calculated Sommerfeld coefficient $\gamma \equiv C_{\rm el}/T$ to the measured electronic specific heat for Nd-LSCO. At $p = 0.27$, $0.36$, and $0.40$, $C_{\rm el} / T$ is found to be constant at low $T$, with $\gamma = 11$, $6.5$ and $5.5 \pm 1$~mJ~/~K$^2$~mol, respectively (see \autoref{fig:choice_of_t}b) \cite{michon2019thermodynamic}. For Nd-LSCO at $p = 0.24$, \citet{michon2019thermodynamic} report a $\log(1/T)$ increase in the specific heat below 10 K. Above 10 K, this increase must terminate to be consistent with the specific heat at $p = 0.27$, as the specific heat generally decreases with increased doping as the band moves away from the van Hove points, and the mass enhancement decreases away from \ps. The difference between the measured specific heat at 6 K and the lower bound set by the $p = 0.27$ data is 20\%. Because the origin of the $\log(1/T)$ increase is unknown, and its effect on the electrical transport is unclear, we take the density of states across our temperature range---from 6 to 25 K---to be constant. We know of at least two cases where $m^{\star}$ is independent of temperature in a metal that does exhibit $T$-linear resistivity. The first case is the electron-doped cuprates, where the mass from quantum oscillations in NCCO at $x = 0.17$ obeys the standard Lifshitz-Kosevich (LK) form, with a constant mass of $m^{\star} = 2.3 m_e$ \cite{helmthesis}. Over the same temperature range, the resistivity of PCCO at $x = 0.17$ is purely $T$-linear \cite{fournier1998insulator}. The second case is Tl-2201 at $p = 0.30$, where quantum oscillations are perfectly LK-like, indicating a constant $m^{\star}$ \cite{bangura_QO_Tl2201}. Over the same temperature range where the quantum oscillations are measured, from 1 to 5 K, the resistivity of Tl-2201 is dominated by the $T$-linear component \cite{proust_tl2201}. Thus there is no clear link between $\log(1/T)$ specific heat and $T$-linear resistivity.

We therefore take a value of $t = 130$~meV as a lower bound on $t$ (see \autoref{fig:choice_of_t}a). The upper bound on $t$ is set by ARPES, because ARPES is not necessarily sensitive to all low-energy renormalizations near the Fermi energy. This upper bound is $t  = 190$~meV \cite{horio2018three}. \autoref{fig:choice_of_t}a shows that a value of $t = 160 \pm 30$~meV---encompassing the lower bound set by specific heat and the upper bound set by ARPES---agrees well with the measured specific heat across the entire doping range, passing through all error bars, from $p = 0.23$ to $p = 0.40$. The fractional reduction in bandwidth from the ARPES value, \textit{i.e.} $t$(ARPES)$/t$($\gamma$), is $1.2$.

\subsection*{Scattering rate models}
In order to eliminate a possible model dependence of the scattering rate to best describe the ADMR data of Nd-LSCO at $p=0.24$, we tested different scattering rate models that we detail below.

\begin{figure}[h!]
\centering
\includegraphics[width=1\columnwidth]{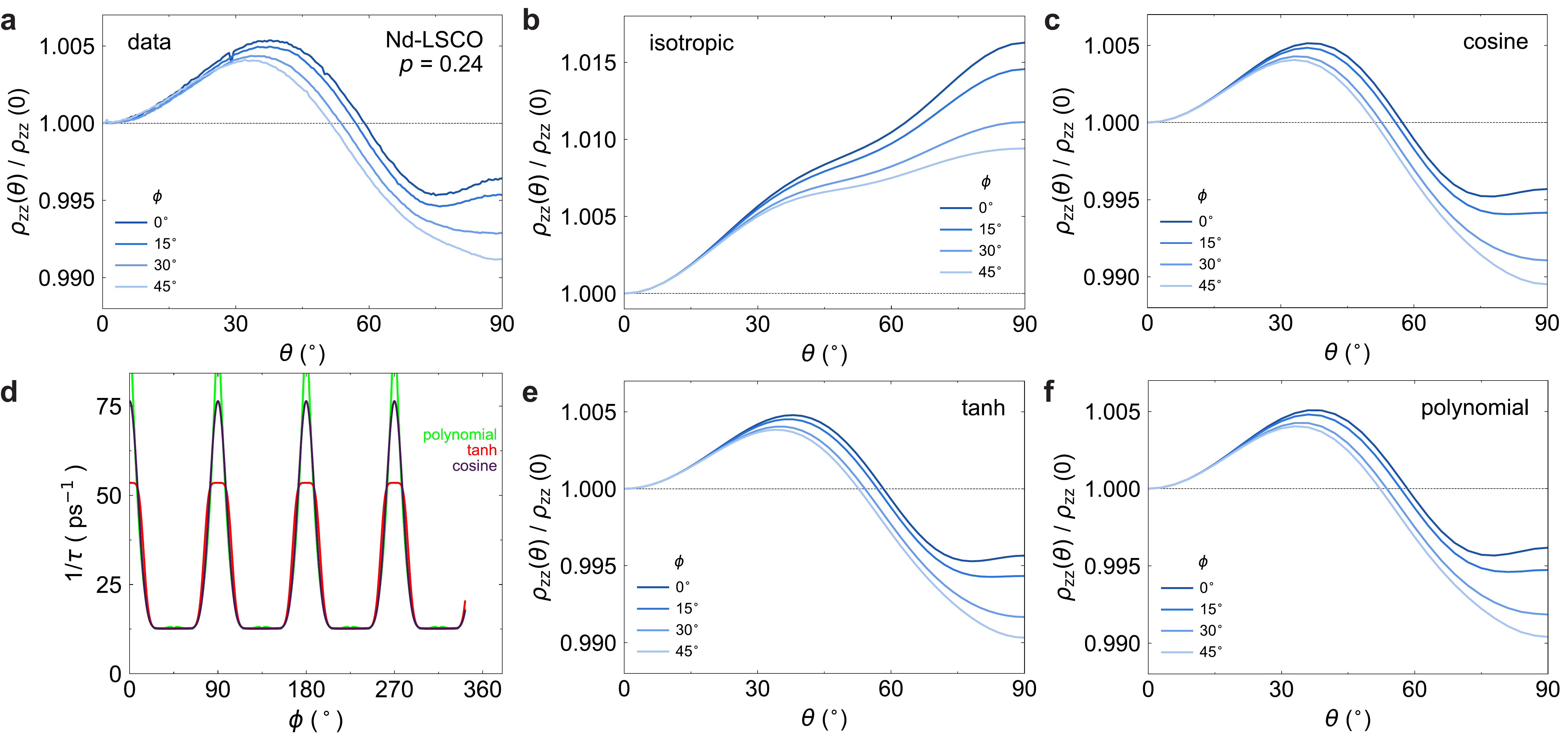}%
\refstepcounter{extended}\label{fig:scattering_models_0p24}
\caption{\textbf{Fit of the Nd-LSCO $\bm{p}$ = 0.24 data with different scattering rate models}. (\textbf{a}) ADMR data on Nd-LSCO $p=0.24$ at $T=25$~K and $B=45$~T; (\textbf{b}, \textbf{c}, \textbf{e}, \textbf{f}) Best fits for the ADMR data in (a) using the Fermi surface in \autoref{fig:phase_diagram}d and, respectively, an isotropic scattering rate model, and three different anisotropic scattering rate models: cosine, $\tanh$ and polynomial; (\textbf{d}) The three different anisotropic scattering rates as a function of the azimuthal angle $\phi$ at $T=25$~K.}
\end{figure}

\textbf{Isotropic scattering rate}. We first consider a constant scattering rate model

\begin{equation}
1/\tau = 1/\tau_{\rm iso},
\label{eq:scattering_isotropic}
\end{equation}

where $1/\tau_{\rm iso}$ is the amplitude of the isotropic scattering rate. With this we try to fit the ADMR data of Nd-LSCO at $p=0.24$. The best fit result is showed in \autoref{fig:scattering_models_0p24}b, which demonstrates that a constant scattering rate model fails to reproduce the data. Instead, the signal increases monotonically out to $\theta = \degr{90}$. The features at $\theta = \degr{40}$ and $\theta = \degr{90}$ are present -- as they reflect the topology of the Fermi surface-- but come with wrong amplitudes and proportions in respect to each other. Using a smaller or higher scattering rate just changes the overall amplitude of the curve, but not the proportions of the features in respect to each other.

\textbf{Anisotropic scattering rate: cosine}. We next consider the most minimalistic anisotropic scattering rate model, one based on a cosine function:
\begin{equation}
1/\tau(\phi) = 1/\tau_{\rm iso}+1/\tau_{\rm aniso}|\cos(2\phi)|^\nu,
\label{eq:scattering_cosine}
\end{equation}
where $1/\tau_{\rm iso}$ is the amplitude of the isotropic scattering rate, $1/\tau_{\rm aniso}$ is the amplitude of the $\phi$-dependent scattering rate, and $\nu$ is an integer. The best fit using this model is plotted in \autoref{fig:admr_scattering}a and \autoref{fig:scattering_models_0p24}c, and parameter values are listed in \autoref{tab:fit_result_0p24_scattering}. The features at $\theta = \degr{40}$ and $\theta = \degr{90}$ are now present with the same amplitudes as the data. With as few parameters as possible, this model captures the trend of the anti-nodal regions of the Fermi surface to have shorter quasiparticle lifetimes in the cuprates \cite{abrahams2000angle,analytis2007angle}, particularly close the van Hove singularity at $p\approx 0.23$. This is the model we used in \autoref{fig:admr_scattering} --- it should be seen as the simplest phenomenological model able to capture the correct shape of the real scattering rate, with the least number of free parameters.

\begin{table}[h!]
  \begin{center}
  \begin{adjustbox}{width=1\textwidth}
    \begin{tabular}{|c|c|c|c|c|c|c|c|c|c|}
    \hline
    $T$ (K)& $1/\tau_{\rm iso}$ (ps$^{-1}$) & $1/\tau_{\rm aniso}$ (ps$^{-1}$) & $\nu$ & $t$ (meV) & $t'$ & $t''$ & $t_z$ & $\mu$ & $p$ \\
    \hline
    25 & $12.595\pm0.002$ & $63.823\pm0.257$ & $12\pm1$ & $160\pm 30$ & $-0.1364t$ & $0.0682t$ & $0.0651t$ & $-0.8243t$ & $0.248$\\
    20 & $11.937\pm0.003$ & $63.565\pm0.759$ & $12\pm1$ & $160\pm 30$ & $-0.1364t$ & $0.0682t$ & $0.0651t$ & $-0.8243t$ & $0.248$\\
    12 & $10.663\pm0.005$ & $63.599\pm0.235$ & $12\pm1$ & $160\pm 30$ & $-0.1364t$ & $0.0682t$ & $0.0651t$ & $-0.8243t$ & $0.248$\\
    6 & $9.628\pm0.049$ & $63.929\pm0.902$ & $12\pm1$ & $160\pm 30$ & $-0.1364t$ & $0.0682t$ & $0.0651t$ & $-0.8243t$ & $0.248$\\
    \hline
    \end{tabular}
    \end{adjustbox}
  \end{center}
 \caption{\textbf{Results of the fit of the Nd-LSCO $\bm{p}$ = 0.24 data with the cosine scattering rate model.} Best fit scattering rate and tight-binding values of the Nd-LSCO $p=0.24$ ADMR data plotted in \autoref{fig:admr_scattering}a. The fit was achieved by the multi-temperature fit procedure described in the above section. Error bars on the scattering rate parameters were obtained following the procedure described in the fitting method section. Error bars on the tight-binding parameters are all $\pm0.0005 t$.}
\label{tab:fit_result_0p24_scattering}
\end{table}

\textbf{Anisotropic scattering rates: tanh and polynomial}. In order to ensure that the cosine model captures the phenomenology of the real scattering rate without being a ``fine-tuned'' model, we now turn to two other anisotropic scattering rate models based on entirely different functions. The first model incorporates a hyperbolic tangent function ( \autoref{eq:scattering_tanh}), the second is a polynomial function in (\autoref{eq:scattering_polynomial}) (the most `adaptive' of the three models).
The ``tanh'' model,
\begin{equation}
1/\tau(\phi) = \frac{1/\tau_{\rm iso}}{|\tanh(a_{\rm 1} + a_{\rm 2} |\cos(2(\phi+\pi/4))|^{a_{\rm 3}})|},
\label{eq:scattering_tanh}
\end{equation}
and the polynomial model,
\begin{equation}
1/\tau(\phi) = 1/\tau_{\rm iso} + |a_{\rm 1}\phi + a_{\rm 2}\phi^2 + a_{\rm 3}\phi^3 + a_{\rm 4}\phi^4 + a_{\rm 5}\phi^5|,~\textnormal{with}~\phi(\textnormal{mod}~\pi/4) \in [0, \pi/4].
\label{eq:scattering_polynomial}
\end{equation}

\begin{figure}[h!]
\centering
\includegraphics[width=0.95\columnwidth]{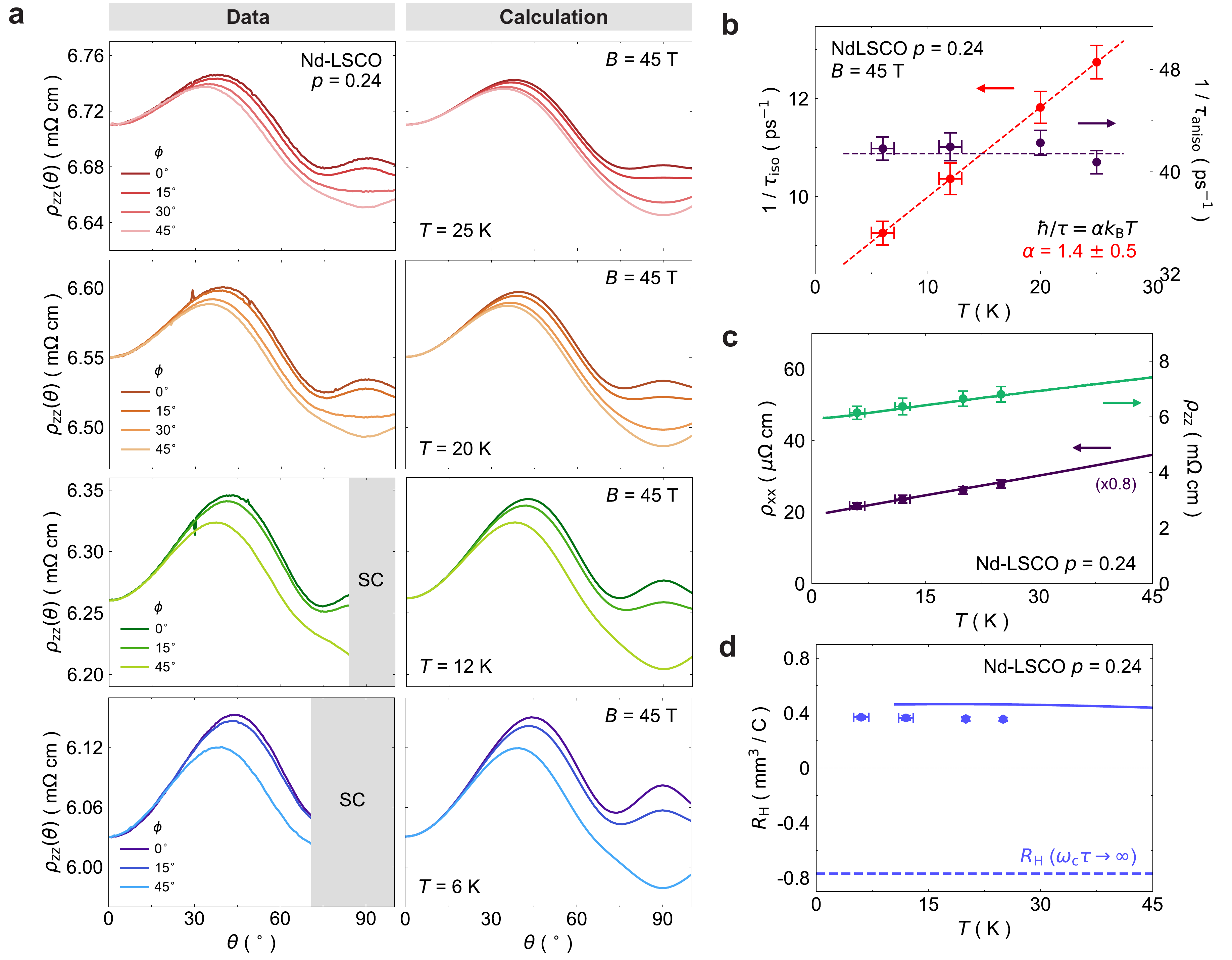}%
\refstepcounter{extended}\label{fig:scattering_tanh}
\caption{\textbf{ADMR and quasiparticle scattering rate of Nd-LSCO at $\bm{p = 0.24}$ for ``tanh'' model.} This figure is the same as \autoref{fig:admr_scattering}a and \autoref{fig:admr_transport}a, b in the main text, except that the ADMR has been fit using the ``tanh'' model instead of the ``cosine'' model (see \autoref{fig:scattering_models_0p24}).
}
\end{figure}

The best fits for these two models are plotted in \autoref{fig:scattering_models_0p24}e and f. The entire temperature dependence and the transport coefficients calculated with the ``tanh'' model are shown in \autoref{fig:scattering_tanh}. The fits are not significantly different from the ``cosine'' model---slightly more refined---which demonstrates that the essential physics is captured by the minimalistic cosine model. \autoref{fig:scattering_models_0p24}d shows that the three anisotropic models all give the same $\phi$-dependence close to the nodes at $\theta=\degr{45}$ and have the same slopes near $\theta=\degr{90}$. The models differ in their absolute values of the scattering rate near $\theta=\degr{90}$: we attribute this small discrepancy to the fact that the scattering rate at $\theta=\degr{90}$ is so high that a small change in curvature in the model can make the value at $\theta=\degr{90}$ vary. Nonetheless, this is just a quantitative difference, as the transport coefficients calculated remain similar, the anisotropic component of the scattering rate remains temperature independent and the isotropic part is $T$-linear for both the ``cosine'' and ``tanh'' models as shown in \autoref{fig:scattering_tanh}b. We do not present the temperature dependence of the ``polynomial'' model because of the long time it takes to converge with many more parameters.

\subsection*{ADMR for B = 35~T}

\begin{figure}[h!]
\centering
\includegraphics[width=0.95\columnwidth]{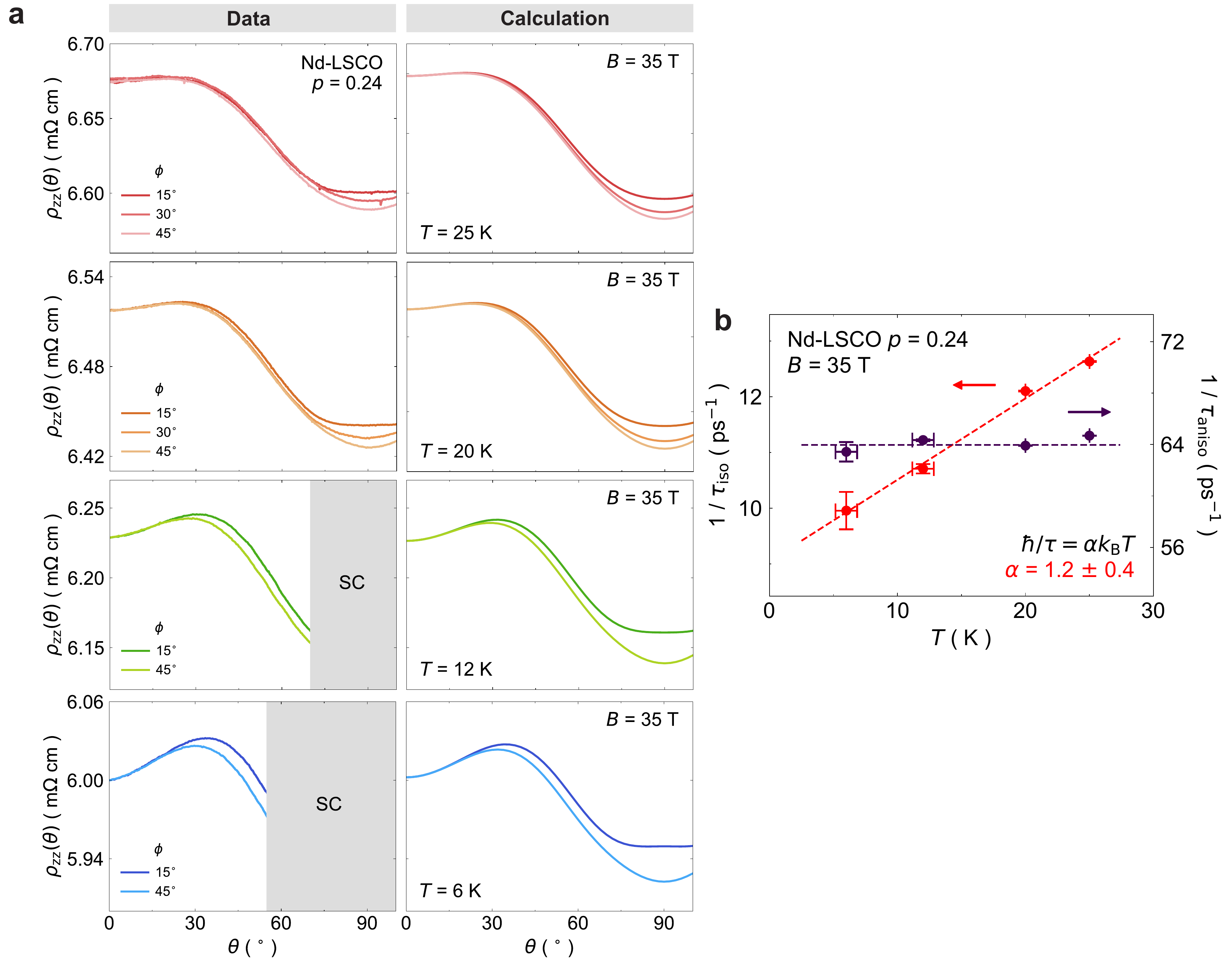}%
\refstepcounter{extended}\label{fig:admr_35}
\caption{\textbf{ADMR and quasiparticle scattering rate of Nd-LSCO at $\bm{p = 0.24}$ for B~=~35~T.} This figure is the same as \autoref{fig:admr_scattering}a and c in the main text except that the ADMR data are taken at $B=35$~T. The fit has been carried out using the ``cosine'' model. Panel b shows that scattering rate values are identical to within a percent of those obtained from the fit to the data at $B=45$~T, shown in figure \autoref{fig:admr_scattering}c.
}
\end{figure}

In \autoref{fig:admr_35}a, we show ADMR data measured at $B=35$~T. Because of the smaller magnetic field, the data are very different from those taken at $B=45$~T – both the magnitude of the ADMR as well as the qualitative features.

By following the same fitting procedure at $B=35$~T, \autoref{fig:admr_35}b we show that we obtain scattering rates and tight-binding parameters that are the same to within 1\% of those that describe the data at $B=45$~T  – the only parameter that changes between the two sets of fits is the magnetic field.

This shows that our scattering rate and tight binding parameters are robust and do not rely on fine-tuning parameters to match the data at $B=45$~T. Moreover, this shows that the scattering rate does not depend on field, and that all magnetoresistance arises from the orbital motion of electrons under the Lorentz force.

\makeatletter
\apptocmd{\thebibliography}{\global\c@NAT@ctr 37\relax}{}{}
\makeatother

\renewcommand\refname{Methods References}

\section*{Data Availability}
Experimental data presented in this paper is available at http://wrap.warwick.ac.uk/152398/. The results of the conductivity simulations are available from the corresponding authors upon reasonable request.

 \section*{Code Availability}
The code used to compute the conductivity is available from the corresponding authors upon reasonable request.

\newpage
\section*{Supplementary Information}

\renewcommand{\thefigure}{S\arabic{figure}}

\subsection*{Tight Binding Model}
We use a three dimensional tight binding model of the Fermi surface that accounts for the body-centered tetragonal crystal structure of Nd-LSCO \cite{horio2018three},
\begin{equation}
  \begin{aligned}
\epsilon(k_x,k_y,k_z)=&-\mu
-2t[\cos(k_xa)+\cos(k_ya)]
\\&-4t'\cos(k_xa)\cos(k_ya)
-2t''[\cos(2k_xa)+\cos(2k_ya)]
\\&-2t_z\cos(k_xa/2)\cos(k_ya/2)\cos(k_zc/2)[\cos(k_xa)-\cos(k_ya)]^2,
  \end{aligned}
	\label{eq:band_structure_0p24}
\end{equation}
where $\mu$ is the chemical potential, $t$, $t'$, and $t''$ are the first, second, and third nearest neighbor hopping parameters, $t_z$ is the inter-layer hopping parameter, $a=3.75$~\AA~is the in-plane lattice constant of Nd-LSCO, and $c/2=6.6$~\AA~is the CuO$_2$ layer spacing. The inter-layer hopping has the form factor $\cos(k_xa/2)\cos(k_ya/2)[\cos(k_xa)-\cos(k_ya)]^2$, which accounts for the offset copper oxide planes between layers of the body-centered tetragonal structure \cite{chakravarty1993interlayer}.

\subsection*{Transport Calculations in a Magnetic Field}
The semi-classical electrical conductivity of a metal can be calculated by solving the Boltzmann transport equation within the relaxation-time approximation. The approach most suitable for calculating angle-dependent magnetoresistance was formulated by \citet{Chambers1952kinetic}. It provides an intuitive prescription for calculating the full conductivity tensor $\sigma_{ij}$ in a magnetic field $\bm{B}$, starting from a tight-binding model of the electronic band structure~$\epsilon(\bm{k})$.
Chambers' solution is
\begin{equation}
\sigma_{ij}  = \frac{e^2}{4 \pi^3}\!\int \!d^3\bm{k} \left(-\frac{df_0}{d\epsilon}\right)v_i\!\left[\bm{k}\!\left(t=0\right)\right]\int^{0}_{-\infty}\!v_j\!\left[\bm{k}\!\left(t\right)\right]e^{t/\tau}dt,
\label{eq:chamb}
\end{equation}
where $\int \!d^3\bm{k}$ is an integral over the entire Brillouin zone, $\left(-\frac{df_0}{d\epsilon}\right)$ is the derivative with respect to energy of the equilibrium Fermi distribution function, $v_i$ is the $i^{\mathrm{th}}$ component of the quasiparticle velocity, and  $\int^{0}_{-\infty}dt$ is an integral over the lifetime, $\tau$, of a quasiparticle.
The Fermi velocity is calculated from the tight binding model as $\bm{v}_{\bm{F}} = \frac{1}{\hbar}\vec{\nabla}_{\bm{k}}\epsilon(\bm{k})$.
The magnetic field, including its orientation with respect to the crystal axes, enters through the Lorentz force, which acts to evolve the momentum $\bm{k}$ of the quasiparticle through $\hbar \frac{d\bm{k}}{\mathrm{d}t} = e \bm{v}\times\bm{B}$. Because the magnetic field is included explicitly in this manner, Chambers' solution has the advantage of being exact to all orders in magnetic field.

The conductivity of a general electronic dispersion relation $\epsilon(\bm{k})$ can be calculated using \autoref{eq:chamb}. The factor $\left(-\frac{df_0}{d\epsilon}\right)$ is approximated as a delta function at the Fermi energy in the limit that the temperature $T$ is much smaller than any of the hopping parameters in $\epsilon(\bm{k})$, as is the case for our experiments. This delta function transforms the integral over the Brillouin zone into an integral over the Fermi surface, and introduces a factor of $1/|\vec{\nabla}_{\bm{k}}\epsilon(\bm{k})|$, which is the density of states. To perform the integrals in \autoref{eq:chamb} numerically, the Fermi surface is discretized, usually into 10 to 15 layers along $k_z$, with 60 to 100 points per $k_z$ layer, and each point is evolved in time using the Lorentz force equation. This moves the quasiparticles along cyclotron orbits around the Fermi surface, and their velocity is recorded at each position and integrated over time. The weighting factor $e^{t/\tau}$ accounts for the scattering of the quasiparticles as they traverse the orbit. In general, $\tau$ is taken to be a function of momentum, $\tau(\bm{k})$, and then the factor $e^{t/\tau}$ is replaced by $e^{\int_{t}^{0} dt'/\tau(\bm k\left(t'\right))}$. \autoref{eq:chamb} can be used to calculate any component of the semiclassical conductivity tensor.
We use it to calculate $\rho_{\rm zz}$, as well as $\rho_{\rm xx}$ and $\rho_{\rm xy}$ in Figure 3 of the main text. Note that because of the highly 2D nature of the Fermi surface of Nd-LSCO, we neglect the off-diagonal components of the conductivity tensor and use $\rho_{\rm zz} \approx 1/\sigma_{\rm zz}$. For $\rho_{\rm xx}$ and $\rho_{\rm xy}$ we invert the full in-plane conductivity tensor.

\subsection*{Cyclotron Frequency}
The product of the cyclotron frequency, $\omega_{\rm c} \equiv \frac{e B}{m^{\star}}$, and the quasiparticle lifetime, $\tau$, is generally seen as a good indicator of whether one should expect to observe quantum oscillations and ADMR. When $\omega_{\rm c} \tau \sim 1$ or greater, quasiparticles complete full cyclotron orbits and the effects of both Landau quantization and Fermi surface geometry are seen in the transport. When $\omega_{\rm c} \tau \ll 1$, on the other hand, quasiparticles scatter too frequently for these effects to be observed. Given that we observe ADMR but not quantum oscillations in these samples, it is worth investigating the structure of $\omega_{\rm c} \tau$ in more detail for Nd-LSCO.

We calculate $\omega_{\rm c}\tau$ for Nd-LSCO $p=0.24$ at $B=45$~T, with $\bm{B}\parallel c$, using the relation
\begin{equation}
\frac{1}{\omega_{\rm c}\tau}=\frac{\hbar}{2\pi eB}\oint\frac{dk}{v_{\perp}(k)\tau(k)},
\label{eq:oct1}
\end{equation}
$v_{\perp}(k)$ is the component of the velocity perpendicular to the field, $1/\tau(k)$ is the total scattering rate, and the line integral follows the closed cyclotron orbit around the total length on the Fermi surface. Using our extracted scattering rate at $T=25$~K for Nd-LSCO $p=0.24$, \autoref{eq:oct1} gives $\omega_{\rm c}\tau = 0.024$ at $\theta = \degr{0}$.

\begin{figure}[h!]
\centering
\includegraphics[width=.9\columnwidth]{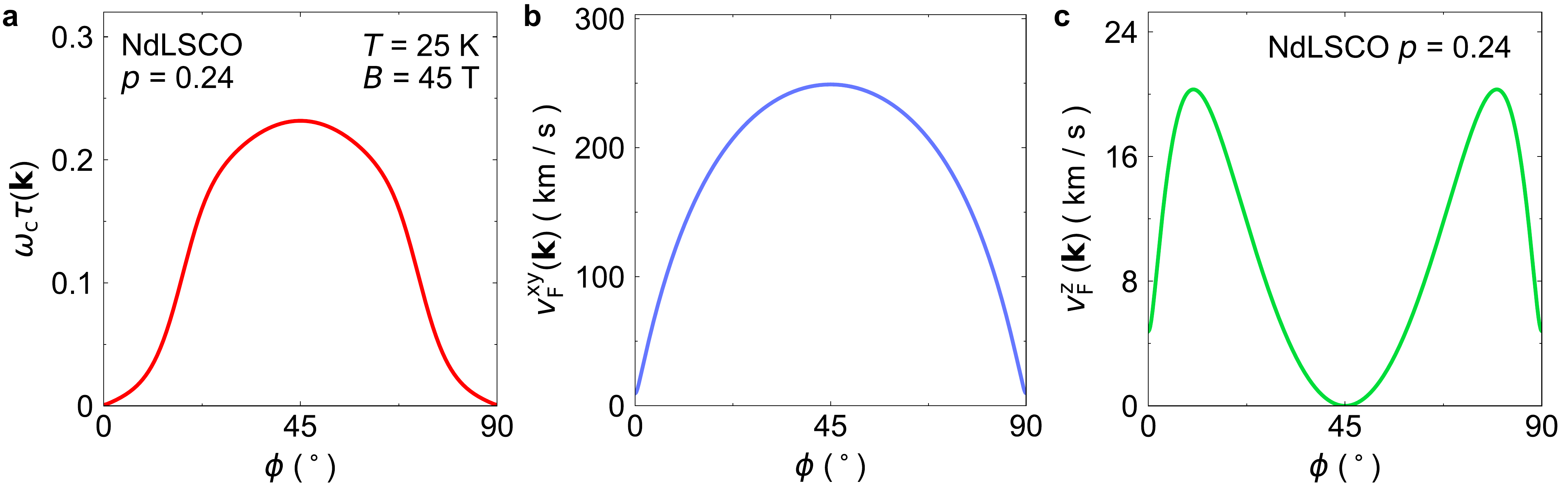}%
\caption{\textbf{$\bm{\omega_{\rm c}\tau}$, velocities around the Fermi surface of Nd-LSCO $\bm{p}$ = 0.24.} (\textbf{a}) Local $\omega_{\rm c}\tau$ as a function of in-plane $\phi$ angle for Nd-LSCO $p=0.24$ at $T=25$~K and $B=45$~T, with $\bm{B}\parallel c$, and closed cyclotron orbit on the $k_z=\pi/c$ Fermi surface; (\textbf{b}) In plane velocity $v_{\rm F}^{\rm xy}$; (\textbf{c}) Out of plane velocity $v_{\rm F}^{\rm z}$, as a function in-plane $\phi$ angle on the $k_z=\pi/c$ Fermi surface of Nd-LSCO $p=0.24$.}%
\label{fig:omegaC}%
\end{figure}

Given this small value of $\omega_{\rm c} \tau$, it may be somewhat surprising that we detect features in the ADMR. Unlike quantum oscillations, however, ADMR does not require quantum coherence around a cyclotron orbit. Instead, ADMR is sensitive to the \textit{local} structure of the FS. Thus the nodal regions of the Fermi surface, with longer quasiparticle lifetimes and smaller cyclotron effective masses, can still contribute to the ADMR even though full cyclotron orbits are prohibited. To visualize this we define a \textit{local} $\omega_{\rm c}\tau$ as a function of each point $\bm k$ around the cyclotron orbit via:
\begin{equation}
\frac{1}{\omega_{\rm c}\tau(\bm k)} = \frac{\hbar}{2\pi eB}\frac{2\pi \bm k}{v_{\perp}(\bm k)\tau(\bm k)} =  \frac{m^*(\bm k)}{eB\tau(\bm k)},
\label{eq:oct2}
\end{equation}
where $m^*(\bm k)=\hbar k/v_{\perp}(\bm k)$ the local effective mass at point $\bm k$.

We parametrize $\bm k$ around a cyclotron orbit by the angle $\phi$, and plot the effective $\omega_{\rm c} \tau$ as a function of $\phi$. \autoref{fig:omegaC} shows that in the nodal region ($\phi=45^\circ$) the effective $\omega_{\rm c}\tau$ is near~$0.3$. This is close to the $\omega_c \tau$ measured in the cuprate Tl-2201, which was found to be $0.45$ \cite{hussey2003coherent}. The $\omega_{\rm c}\tau$ integrated around a full cyclotron orbit, on the other hand, is $0.024$, explaining why quantum oscillations (which require full cyclotron orbits) are not visible in Nd-LSCO.

\subsection*{ADMR vs ARPES Elastic Scattering}

\begin{figure}[h!]
\centering
\includegraphics[width=.7\columnwidth]{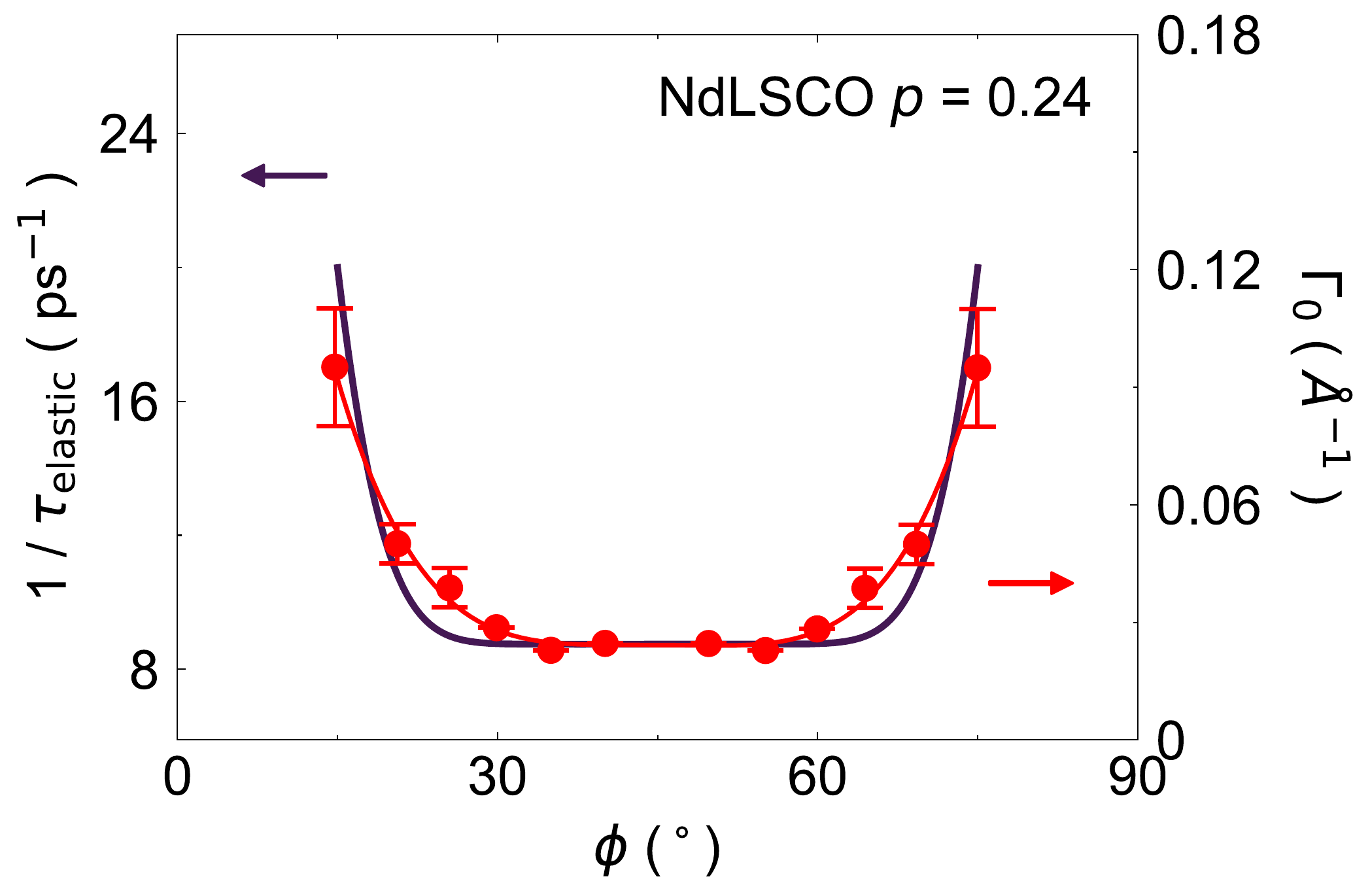}%
\caption{\textbf{Comparison between ADMR scattering rates and ARPES linewidths} | The red points are the linewidths---proportional to the single-particle scattering rate---measured by ARPES in LSCO at $p=0.23$, taken from figure S2 of \citet{chang2013anisotropic}. The purple curve is the elastic part of the scattering rate we obtain by fitting the ADMR for Nd-LSCO at $p=0.24$.}%
\label{fig:ADMR_vs_ARPES_scattering}%
\end{figure}

\subsection*{Origin of the Linear Magnetoresistance}

The following is a simple example of how $B-$linear magnetoresistance arises naturally from Fermi surface and scattering-rate anisotropy. Semiclassical magnetoresistance generically has two regimes: the low-field regime, where $\rho \propto B^2$, and the high field regime, where $\rho$ saturates and is $B$-independent \cite{abrikosov2017fundamentals}. The crossover between $B^2$ and $B$-independent naturally produces a region of $B$-linear resistivity in between \cite{peierls1931theorie}. The range in magnetic field over which the magnetoresistance appears to be $B$-linear depends on the microscopic details of the Fermi surface and the scattering rate. In particular, Fermi surface with large anisotropy in $v_F$ or $\tau$ (or both) will have a large region of $B$-linear magnetoresistance.

To illustrate how $B$-linear magnetoresistance arises from \autoref{eq:chamb}, consider a square Fermi surface (so that $v_i$ is constant on each side) with an isotropic $\tau$ everywhere except for the corners of the square where $\tau$ is zero. Apply a magnetic field perpendicular to the plane of the square such that quasiparticles traverse the sides of the square and then hit the corners and scatter immediately. 

The contribution to the conductivity from a single quasiparticle (integrating over the entire Fermi surface will only introduce an overall numerical factor, as $v_x$ is a constant for 2 of the sides and 0 for the other 2) is computed from \autoref{eq:chamb} as $\sigma_{xx} \sim v_x^2 \int^{\infty}_{0} e^{-t/\tau} dt$, where we have used the fact that $v_x$ is a constant and we have switched the direction of $t$ for greater clarity in this example (with no loss of generality.) The time taken for a quasiparticle to traverse a side of the square is given by solving the Lortenz force equation and is $\Delta t = \hbar \Delta k / (e v_x B)$, where $\Delta k$ is the distance in momentum space from where the quasiparticle started to the corner it will encounter first (which depends on the sign of the magnetic field). Once the quasiparticle reaches the corner, $\tau$ becomes zero and the integral is cut off. The conductivity is then $\sigma_{xx} \sim v_x^2 \tau(1-e^{\frac{-\hbar \Delta k}{e v_x B \tau}})$. In the limit that $B$ is large this expression reduces to $\sigma_{xx} \propto 1/B$, resulting in $B$-linear $\rho_{xx}$ ($\sigma_{xy}$ is zero in this model because quasiparticles never make it past a corner where constant $v_x$ changes to constant $v_y$).

As the lifetime $\tau$ is increased from zero at the corners, there will be an offset to the conductivity resulting in $B^2$ resistivity at low-fields and $B$-linear resistivity at high fields. Only once $\omega_c \tau>>1$ is reached will the resistivity saturate---this can be an arbitrarily high field scale if $\tau$ is made arbitrarily small at the corners. A more realistic Fermi surface and scattering rate will change this calculation in quantitative ways but the three regimes remain, and with a very anisotropic $\tau$ (or $v_F$) one can have a very large region of $B$-linear resistivity.




\end{document}